\documentclass[twocolumn,secnumarabic,amssymb, nobibnotes, aps, prd]{revtex4-2}

\newcommand{\nnm}{\nonumber}

\usepackage{color}
\usepackage{graphicx}
\usepackage{epsfig}
\begin{document}

\title{Five and three quantum dot systems as apparatuses for measuring energy-levels }%

\author{Tetsufumi Tanamoto}
\affiliation{Department of Information and Electronic Engineering, Teikyo University,
Toyosatodai, Utsunomiya 320-8511, Japan} 
\email{tanamoto@ics.teikyo-u.ac.jp}

\author{Tomosuke Aono}
\affiliation{Department of Electrical and Electronic Systems Engineering, Faculty of Engineering, Ibaraki University, Hitachi 316-8511, Japan} 




\begin{abstract}
A quantum dot (QD) system provides various quantum physics of nanostructures.
So far, many types of semiconductor QD structures 
have been fabricated and investigated experimentally and analyzed theoretically.
Presently, QD systems have attracted considerable attention as 
units for the qubit system of quantum computers.
Therefore, it is vital to integrate QD systems as measurement 
devices in addition to qubits.
Here, we theoretically investigate the side-QD system as a measurement apparatus 
for energy-levels of the target QDs.
We formulate the transport properties of both three and five QDs based on the Green functions method.
The effects of the energy-difference of two side-QDs on the measurement current are calculated.
The trade-off between the strength of the measurement and the back-action induced by the measurement is discussed.
It is found that the medium coupling strength is appropriate for reading out the difference of the 
two energy-levels.
\end{abstract}

\maketitle

\section{Introduction}

Quantum dot (QD) systems have been providing various topics in quantum physics for electronic systems.
The interference between QDs and channel electrons is an important 
phenomenon that characterizes the transport properties of the system.
The great developments of semiconductor nanofabrication processes enable experimentalists to directly observe the
nanoworld by using the abundant technologies of the 
miniaturization of semiconductor devices.
Numerous excellent experimental works have been carried out in this mesoscopic field~\cite{Buks,Fujisawa1,Fujisawa2,Tarucha,Kobayashi2,Clerk,Johnson,Kroner,Sasaki,Rushforth}.
Recently, many QD systems have become a target structure of spin qubits 
because spin qubits enter into their development phase with many QDs~\cite{Hensgens,Mills,Fedele,Ha}.
Thus, the transport properties of many QDs are of new-found interest in several fields of physics and engineering.

In QD systems, the changes in energy-levels of QDs to external controls 
are very small, and detecting energy-levels is very difficult~\cite{Clerk2,Zurek,Bethke,Kodera,Pan}. 
Generally speaking, we can obtain the knowledge of the energy-levels indirectly 
through the current line attached close to the target QDs.
In addition, we have to consider the effect of the back-action by the measurements. 
In order to obtain strong signals, the coupling between the target structure and 
the measurement structure should be large. 
However, the strong coupling to the target structure tends to 
destroy the coherence of the system. 
The trade-off between the measurement and the back-action is an important issue.

In this study, we theoretically describe how to measure the difference between 
the energy-levels of two QDs by using the side-QDs system.
We focus on the measurement of the QDs in the side-QD system, as shown in Fig. 1.
In the conventional side-QD structure (Fig.~1(a)), 
the arrangement of QDs is symmetric to the center current line (S3-QD3-D3 in Fig.~1(a)),
therefore, we cannot judge which of the QDs has a higher energy-level 
when we measure only the current of the structure in Fig.~1(a).
However, we can distinguish the two QDs by adding two other current lines, 
as shown in Fig. 1(b). 
For distinguishing the two energy-levels, it is sufficient to compare the currents separately. 
For example, by switching on the current line 1 while the other two current lines 
are switched off, the current line 1 reflects only the energy-levels of the QD2.
By combining this with the case where only the current line 3 is switched on, 
we will be able to judge which of the QD2 and QD4 has a higher energy-level.
Similarly, we can use the case where the current line 5 is switched on while the other two currents are switched off. 
On the other hand, when the three currents flow at the same time, 
we can consider an interesting process that does not appear for the separate current detection.
When the three current lines are simultaneously switched on,
new current passes are generated from the source S$i$ to the drain D$j$ ($i\neq j$) 
through the QD between the two current lines.
It is expected that these passes enhance both the measurement and the back-action.
We numerically calculate the transport properties of the Figs. 1(a) and (b), 
and discuss the trade-off of the coupling strength and the back-action.
Hereafter, we call Figs.~1(a) and (b) as the 'three-QD' and 'five-QD' cases, respectively.

The side-QD structures have been mainly investigated as the typical setup for observing the Fano effect,
in which the current shows a dip via the interference between the energy-level of the QD 
and the channel current~\cite{Kobayashi2,Clerk,Johnson,Kroner,Sasaki,Rushforth,Kim,Kang,Durganandini,Cornaglia,Aono1,Aono2,Tana1}.
Moreover, the side-QD structures with two QDs have been called 
the two-impurity Kondo effects.
In the early research, the energy-levels of the two QDs were the same~\cite{Izumida,Coleman,Coleman1,Read}, 
and recently the difference of the two energy-levels is treated to be more widely.
In~\cite{Shahbazyan94}, the resonant tunneling effect through the two impurity levels was discussed, 
and it was found that a significantly narrowed peak structure superimposes over a broad peak structure because of the coupling between the levels and the electrode. 
In addition, the conductance is sensitively affected by the difference in the impure energy-levels.
These results are analogous to the Dicke effect in quantum optics~\cite{Dicke53}, 
where fast and slow relaxation modes appear owing to the interaction between the coupled relaxation channels. 
Similar effects have been extensively discussed for electrical conduction in mesoscopic systems~\cite{Brandes05}. 
In two-side QD systems, the Dicke-like effect has been discussed in terms of the Kondo effect~\cite{Orellana1,Ulloa,Wang,Domanski,Orellana2,Trocha}. 

The structure of many QDs with many current lines will be required in 
the integration of the semiconductor qubits.
This is because packing qubits and the detection current line into a small area 
will be important to maintain the decoherence time of the system.
The detection of the energy-difference between the two QDs are required in many cases of quantum computing systems.
The first example is the detection of the gradient magnetic fields~\cite{Takeda,Struck,Tana1},
which is important to control the qubits individually.
The second example is the detection of two qubits 
in the FinFET structure~\cite{Lansbergen,TanaAIP}.
In~\cite{TanaAIP}, QDs embedded between the channel of the FinFET work as the qubits. 
The results of the final qubit state affect the energy-levels. 
Thus, we aim to study how the current characteristics of the channel
reflect the difference of the energy-levels of two QDs.
The three current lines of Fig. 1(b) is the simple case 
of the integration of the qubits and measurement apparatus.

\begin{figure}
\centering
\includegraphics[width=8.0cm]{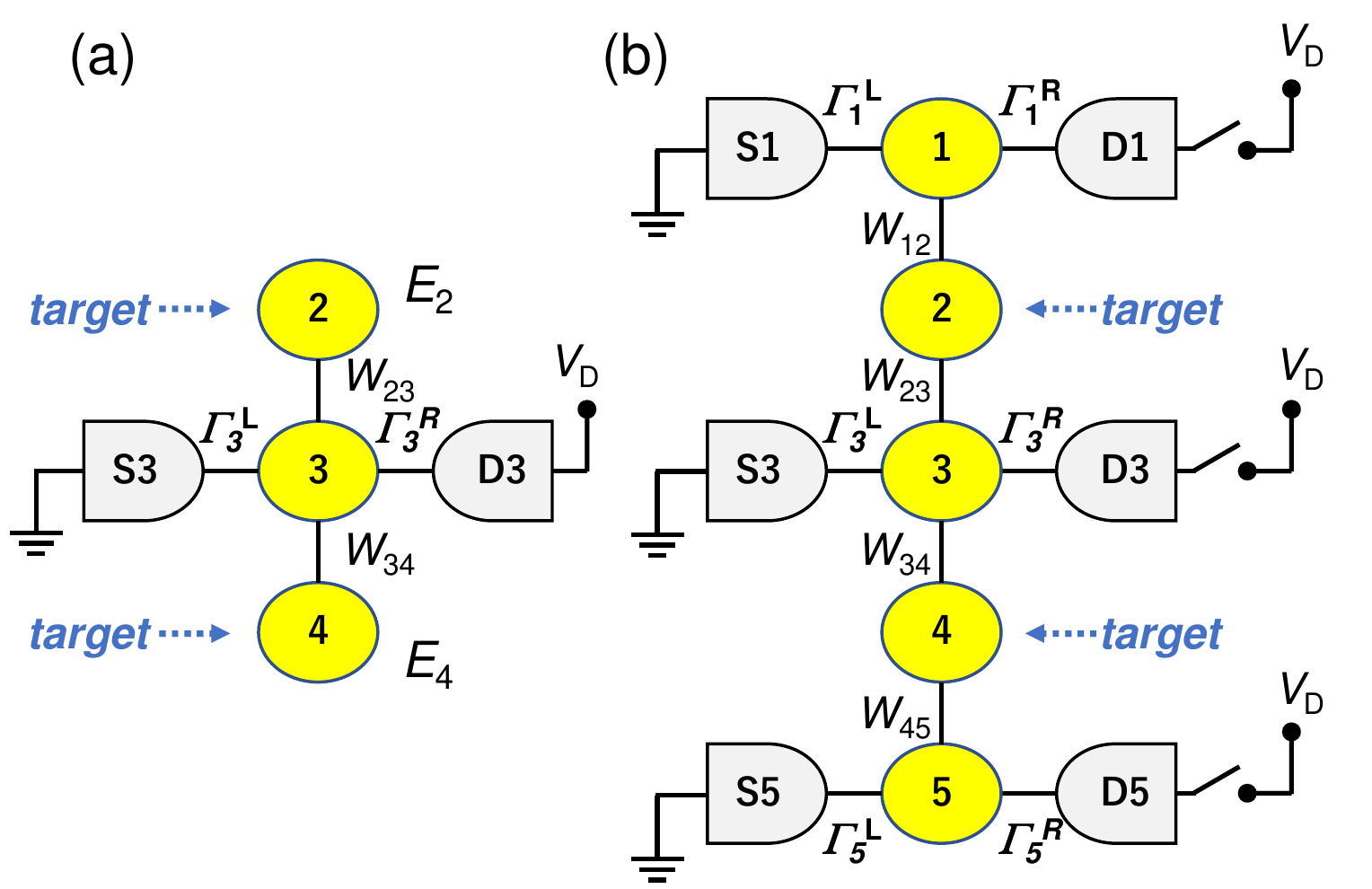}
\caption{Quantum dot (QD) system considered in this study.
In each figure, the center circles show the QDs, 
and the voltages are applied between the sources S$i$ and the drains D$i$ ($i=1,3,5$).
(a) Three QDs system, which includes two side-QDs and one current line.
(b) Five QDs system, which includes the three current lines. 
$\Gamma_i^\alpha$ and $W_{ij}$ are the tunneling couplings between the electrodes ($\alpha=L,R$) and the QDs
and those between two QDs, respectively ($i=1,3,5$, $j=2,4$). 
This study targets the detection of the energy-levels of QD2 and QD4.
The right-hand sides of the current lines in Fig. 1(b)  can be 
switched on and off to select the currents.
By choosing the current lines, the relative energy-level 
between QD2 and QD4 can be detected.
}
\label{fig1}
\end{figure}


We use the Green function methods developed by \cite{Meir1,Jauho},
which enable us to formulate the current characteristics.
The formulation of the five QD system is very complicated, 
and therefore, it is better to observe the characteristics of the system 
without the Kondo effect.
Moreover, it seems that it is not easy to experimentally observe the two-channel Kondo~\cite{Potok,Zhu}. 
In this study, we neglect the Kondo effect and on-site Coulomb interaction in each QD.

The rest of this study is organized as follows.
In Section~\ref{sec:green}, we show our formalism using 
the standard Green function method. 
In Section~\ref{sec:circuit}, we explain our measurement setup.
In Section~\ref{sec:results}, we show the numerical results 
of our method.
In Section~\ref{sec:discussion}, we discuss our results.
In Section~\ref{sec:conclusion}, we summarize and conclude this study.

\section{Green function methods}\label{sec:green}
We investigate the transport properties of 
both the three and five QD systems depicted in Fig.~\ref{fig1}.
The formulation of the three QD systems is the case $W_{12}=0=W_{45}$ of the 
five QD system. Thus, we derive the formula of the five QD system.
The Hamiltonian of the five QD system is given by
\begin{eqnarray}
H&=&\sum_s \sum_{i=1}^5 E_i d_{is}^\dagger d_{is}  
\!+\! \sum_s\sum_{i=1,3,5}\!\!\sum_{\alpha=L,R}\sum_{k_\alpha} E_{k_\alpha} c_{ik_\alpha,s}^\dagger c_{ik_\alpha,s}
\nonumber \\ 
&+&\sum_s \sum_{i=1,3,5}\!\sum_{\alpha=L,R}\sum_{k_\alpha} 
[V_{k_\alpha,s,i} c_{ik_\alpha,s}^\dagger d_{is}+V_{k_\alpha,s,i}^* d_{is}^\dagger c_{ik_\alpha,s} ]
\nonumber \\
&+& \sum_{i=1}^{4}\sum_s W_{i,i+1}(d_{is}^\dagger d_{i+1,s}+h.c.),
\label{Hamilt}
\end{eqnarray}
where $c_{ik_s}^\dagger$ ($c_{ik_s}$) creates (annihilates) an electron with momentum $k$ 
and spin $s$ in the $i$-leads ($i=1,3,5$), and $d^\dagger_{is}$ ($d_{is}$) creates (annihilates) an electron in the QDs ($i=1,...,5$).
We assume that there is one energy-level in each QD.

Following~\cite{Meir1,Jauho}, the current $I_{iL}$ of the $i$-th left electrode 
is derived from the time derivative of the number of electrons $N_{iL}\equiv \sum_{k_Ls}c_{ik_Ls}^\dagger c_{ik_Ls}$ 
by the left electrode, given by
\begin{eqnarray}
I_{iL}(t)&=&-e\left\langle \frac{d N_{iL}}{dt}\right\rangle =-\frac{ie}{\hbar} \langle [H,N_{iL} ]\rangle, \nnm \\
&=&\frac{ie}{\hbar} \sum_{k_L,s}[V_{k_Ls,i}\langle c_{ik_L s}^\dagger d_{is} \rangle
-V_{k_Ls,i}\langle d_{is}^\dagger c_{ik_Ls} \rangle] \nnm \\
&=& \frac{2e}{\hbar} {\rm Re} \left\{
\sum_{k_Ls} V_{k_L s,i} G^<_{d_{is},c_{ik_Ls}}(t,t)
\right\}\nnm\\
&=& \frac{2e}{\hbar} \int dE {\rm Re} \left\{
\sum_{k_Ls} V_{k_Ls,i}G^<_{d_{is},c_{ik_L s}}(E)
\right\}, \label{currentformula}
\end{eqnarray}
where 
\begin{eqnarray}
G^<_{d_{is},c_{ik_\alpha s}}(t,t')&\equiv& i \langle c_{ik_\alpha s}^\dagger(t') d_{is}(t)\rangle, \label{Gfundk} \\
G^<_{c_{ik_\alpha s},d_{is}}(t,t')&\equiv& i \langle d_{is}^\dagger(t') c_{ik_\alpha s}(t)\rangle, 
\end{eqnarray}
and
\begin{eqnarray}
G^<_{c_{ik_\alpha s},d_{is}}(t,t)=-[G^<_{d_{is},c_{ik_\alpha s}}(t,t)]^*.
\end{eqnarray}
Assuming $I=I_L=-I_R$, we symmetrize the current, 
$I=(I_L+I_L)/2=(I_L-I_R)/2$ in the following.
Hereafter, we assume that the spin-flip process is neglected, and the suffix $s$ is omitted.

The Green functions are derived using the equation of motion method~\cite{Jauho}. 
For example, the time-dependent behavior of the 
operator $d_{i}(t)$ is derived from 
$i\hbar \frac{d~d_{i}(t)}{dt}=[H,d_{i}(t)]$, and we have
\begin{eqnarray}
\omega d_{i}(\omega) = [H, d_{i}(\omega)].
\label{eqmotion}
\end{eqnarray}
As shown in the Appendix, by combining various 
pairs of the operators, all Green functions are obtained.

The Green functions of the electrodes ($\alpha=L,R$) are the free-particle Green functions given by
\begin{eqnarray}
g_\alpha ^<(k,\omega)&=& 2\pi i f (E_{k_\alpha}) \delta(\omega-E_{k_\alpha}), \\
g_\alpha ^>(k,\omega)&=& 2\pi i (f(E_{k_\alpha})-1) \delta(\omega-E_{k_\alpha}), \\
g_\alpha^r(k,\omega)&=& \frac{1}{\omega-E_{k_\alpha} +i\delta},
\end{eqnarray}
where $f(\epsilon)$ is the Fermi distribution function.
The Green functions of the QDs are given by
\begin{eqnarray}
g_{di}^r (\omega)
&=&1/[\omega-E_{i} -\Sigma_i^r(\omega) ] \nnm\\
&=&\frac{\omega-E_{i} -\Lambda_i(\omega)- \frac{i}{2} \Gamma_i}
       {[\omega-E_{i} -\Lambda_i(\omega)]^2+ \Gamma_i^2/4} 
=\frac{b_i(\omega)^*}
        {D_i(\omega)}, \\
b_i(\omega)&\equiv&\omega-E_{i} -\Lambda_i(\omega) +i\Gamma_i/2,       \\
D_i(\omega) &\equiv &[\omega-E_{i} -\Lambda_i(\omega)]^2+ \Gamma_i^2/4,   \\       
\Sigma_i^{r,a}(\omega) &=& \sum_{\alpha}\sum_{k_\alpha} \frac{|V_{k_\alpha i}|^2}{\omega -E_{k_\alpha} \pm i\eta} 
=\Lambda_i(\omega) \mp \frac{i}{2} \Gamma_i, \\
\Gamma_i &\equiv& \Gamma_i^L  +\Gamma_i^R, \ \
\Lambda_i(\omega)  \equiv  \sum_{\alpha}\sum_{k_\alpha}  \frac{|V_{k_\alpha i}|^2}{\omega -E_{k_\alpha}},
\\
g_{di}^< (\omega)&=&g_{di}^r(\omega) \Sigma_i^< (\omega) g_{di}^a (\omega) 
                 =\frac{i [\Gamma_i^Lf_{iL}(\omega)+\Gamma_i^R f_{iR}(\omega)]}
                 {[\omega-E_i-\Lambda_i(\omega)]^2+ \Gamma_i^2/4}, \nnm \\ 
\end{eqnarray}
where $\Lambda_i(\omega)$ is assumed to be constant and included in $E_i$ in the following 
($\Lambda_2=\Lambda_4=0$, $\Gamma_2=\Gamma_4=\delta$).
The $f_{i\alpha}(\omega)$ is the Fermi distribution function given by
$f_{i\alpha}(\omega)=[\exp[(\omega-\mu_\alpha)/(k_BT)]+1]^{-1}$ ($k_B$, $\mu_\alpha$ and $T$ are 
the Boltzmann constant, the chemical potential of the $\alpha$-electrode, and the temperature).
The coupling coefficients of the leads are given by
\begin{equation}
\Gamma_i^{\alpha}(\omega)=2\pi \sum_{k_\alpha} |V_{k_\alpha}^{(i)}|^2\delta (\omega-E_{k_\alpha}).
\end{equation}
After the long derivation process, the retarded and advanced Green functions $G_{dij}$ ($r$ and $a$ are omitted) are given by
\begin{eqnarray}
\!\!\! G_{d11}(\omega)\! &=&\!\! \frac{[1\!-\! C_{32}(\omega)][1\!-\! C_{54}(\omega)]\!-\! \!C_{34}(\omega)}{\Delta_c}
g_{d1}(\omega), \nnm \\ \label{ggd11}\\
\!\!\! G_{d31}(\omega)\!\! &=&\!\! \frac{[1-C_{54}(\omega)]}{\Delta_c} W_{12}W_{23}C_{32}(\omega)g_{d1}(\omega),\\
\!\!\! G_{d51}(\omega)\!\! &=&\!\! \frac{W_{12}W_{23}C_{32}(\omega) W_{43}W_{54}C_{54}(\omega)}{\Delta_c}\! g_{d1}(\omega)\,\\
\!\!\! G_{d13}(\omega)\!\! &=&\!\! \frac{[1-C_{54}(\omega)]}{\Delta_c} W_{23}W_{12}C_{12}(\omega) g_{d3}(\omega), \\
\!\!\! G_{d33}(\omega)\!\! &=&\!\! \frac{[1-C_{54}(\omega)][1-C_{12}(\omega)]}{\Delta_c}g_{d3}(\omega),\\
\!\!\! G_{d53}(\omega)\!\! &=&\!\! \frac{[1-C_{12}(\omega)]}{\Delta_c} W_{43}W_{54}C_{54}(\omega) g_{d3}(\omega),  \\
\!\!\! G_{d15}(\omega)\!\! &=&\!\! \frac{W_{45}W_{34}C_{34}(\omega)W_{23}W_{12}C_{12}(\omega)}{\Delta_c}\! g_{d5}(\omega),  \\
\!\!\! G_{d35}(\omega)\!\! &=&\!\! \frac{W_{45}W_{34}C_{34}(1-C_{12}(\omega))}{\Delta_c} g_{d5}(\omega), \\
\!\!\! G_{d55}(\omega)\!\! &=&\!\! \frac{[1\!\!-\!\! C_{34}(\omega)][1\!-\!\! C_{12}(\omega)]\!-\!\! C_{32}(\omega) }{\Delta_c} 
g_{d5}(\omega),\label{ggd55}
\end{eqnarray}
where 
\begin{eqnarray}
C_{ij}&\equiv &|W_{ij}|^2 g_{di}g_{dj},
\\
\Delta_c&\equiv &
 [1-C_{54}](1-C_{12}-C_{32})-C_{34}+C_{12}C_{34}. 
\end{eqnarray}

\begin{figure}
\centering
\includegraphics[width=8.5cm]{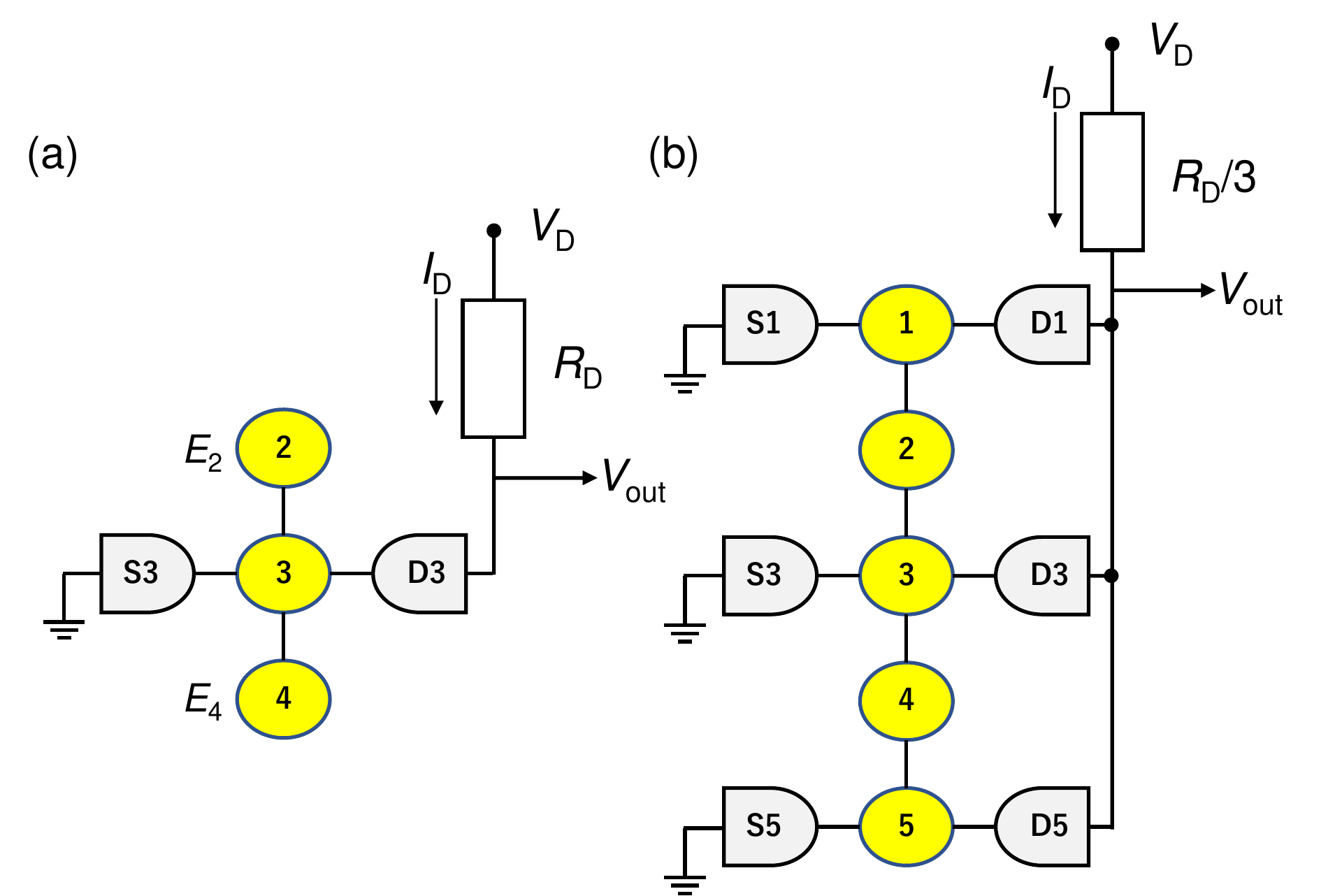}
\caption{Simple setup of converting the current change of the QD system into voltage change
for both (a) the three-QD case and (b) the five-QD case.
We take the resistance $R_D$ as the same order of the QD system, such as $R_D=29.4$k$\Omega$.}
\label{figamp}
\end{figure}

In addition, when three Green functions $A$, $B$, and $C$ have the relation 
of $A=BC$, 
the lesser Green function $A^<$ can be derived from~\cite{Meir1,Jauho} 
\begin{eqnarray}
A^<(E)=B^r(E)C^<(E)+B^<(E)C^a(E).
\label{lesser}
\end{eqnarray}
The derivation of the lesser Green functions is more complicated 
than that of the retarded and advanced Green functions.
After the long derivation process using Eq.(\ref{lesser}), we have the current formula for the five QDs, 
given by (see Appendix)
\begin{widetext}
\begin{eqnarray}
I&=& \frac{e}{\hbar} \int \frac{d\omega}{2\pi} \left\{
\frac{|W_{12}W_{23}W_{34}W_{54}|^2}{|\Delta_c^r|^2}
\frac{1}{D_1D_2D_3D_4D_5} F_{12345} 
+\frac{ |1-C_{12}^r|^2 }{|\Delta_c^r|^2}\frac{|W_{34}W_{54}|^2}{D_3D_4D_5} F_{345} 
+\frac{ |1-C_{54}^r|^2 }{|\Delta_c^r|^2}\frac{|W_{23}W_{12}|^2}{D_1D_2D_3} F_{123} 
\right.\nnm \\
&+&   
\left.
  \frac{|(1-C_{34}^r)(1-C_{12}^r) -C_{32}^r|^2 }{|\Delta_c^r|^2 D_5}F_{55}(\omega) 
+ \frac{|(1-C_{32}^r)(1-C_{54}^r) -C_{34}^r|^2 }{|\Delta_c^r|^2 D_1}F_{11}(\omega)
+\frac{|1-C_{54}^r|^2|1-C_{12}^r|^2 }{|\Delta_c^r|^2D_3}
F_{33} (\omega)\right\},  \label{current5}
\end{eqnarray}
\end{widetext}
where $F_{ii}$, $F_{12345}$, $F_{123}$, and $F_{345}$ are defined in Appendix.
When $W_{12}=W_{45}=0$ in Eq.(\ref{current5}),
we have the current of the three QDs (Fig.1(a)), given as 
\begin{eqnarray}
I_3 \!
&=& \! \frac{e}{\hbar}\int \frac{d\omega}{2\pi}
 \frac{\Gamma^L_3\Gamma^R_3 }{|1-C_{32}-C_{34}|^2D_3}
[f_{3L}(\omega)-f_{3R} (\omega)]. \ 
\label{J3QD}
\end{eqnarray}
The density of the states (DOS) is calculated by
\begin{equation}
\rho(\epsilon)= -\frac{1}{\pi} {\rm Im} [G_{d11}^r+G_{d33}^r+G_{d55}^r](\epsilon).
\label{eqDOS}
\end{equation}
In the following, we mainly show the results of $T\rightarrow0$ limit, 
where $\partial f(\epsilon) /\partial \epsilon \rightarrow -\delta(\epsilon -E_F)$.

\section{Circuit detection}\label{sec:circuit}
We would like to detect the difference of $E_2$ and $E_4$ by using a simple circuit.
In a conventional circuit, the voltage signal is better for output than the current signal. 
In order to transform the current change into voltage change, 
the additional resistor $R_D$ is set to the drain part of the QD system. 
Here, we consider a simple measurement system, as shown in Fig.\ref{figamp},
where Ohm's law leads to the following relation
\begin{equation}
V_{\rm D}=I_D R_D+ V_{\rm out}
\end{equation}
The current $I_D$ is the function of the 
$V_{\rm out}$; thus, this equation should be solved self-consistently.
However, by assuming that the applied voltages are low, and using $I_D=\sigma V_{\rm out}$, we have
\begin{equation}
V_{\rm out}=\frac{1}{1+\sigma R_D} V_{\rm D} 
\end{equation}
In order to effectively reflect the change in $\sigma$, 
the resistor should be in the order of $\sigma^{-1}$ such that 
\begin{equation}
R_D = 1/\sigma \sim \hbar/(2e^2)
\end{equation} 
The amplifying rate is given by
\begin{equation}
\frac{dV_{\rm out}}{d[E_2-E_4]}=-\frac{R_D}{(1+\sigma R_D)^2} \frac{d\sigma}{d[E_2-E_4]}V_{\rm D}. 
\end{equation}
In Fig.~1(a), we take $\sigma=\sigma_3$, and in Fig. 1(b), we take $\sigma=\sum_{i=1,3,5} \sigma_i$,
where $\sigma_i$ is the conductance of the $i$-th current line. 
The relation between the conductance $\sigma_i$ and the transmission coefficient $\mathcal{T}_i$ is given by
\begin{equation}
\sigma_i=\frac{2e^2}{h}\mathcal{T}_i. \label{trans}
\end{equation}
The shot noise is simply estimated by~\cite{Kobayashi,Blanter}
\begin{equation}
S=\frac{2 e^2}{\pi \hbar} \sum_i \mathcal{T}_i(1-\mathcal{T}_i) |eV|,
\end{equation}
where the sum is taken for $i=1,3,5$ for the five-QD case and only $i=3$ for the three-QD case, respectively.
The measurement time is defined by~\cite{Schon}
\begin{equation}
t_{\rm meas}^{-1}\equiv \frac{(\Delta I)^2}{4S_I},
\label{eqmeasurement}
\end{equation}
where we take $\Delta I$ as the difference of the current from that at $E_2=E_4$.
Moreover, we exclude the region of $\Delta I=0$ in the numerical results below.
In the calculation of the noise power $S$, we need the concrete value for 
the applied voltage $V_D$. 
For $V_D=10\sim 100 \mu$eV~\cite{Buks,Fujisawa1}, 
the current $I$ is in order of $I \sim V_D/[2R_D]\sim$ 0.17-1.7 nA, 
where approximately 10-100 electrons flow per 1 ps.
Here, we assume $V_{\rm D}$=1meV~\cite{Fujisawa2}.
Regarding the values for the $\Gamma$, 
$\Gamma=0.5\mu$eV is used in~\cite{Buks}, and 
$\Gamma=$(3ns)$^{-1}$ is used in~\cite{Fujisawa1,Fujisawa2}. 
Here, we take $\Gamma_0=10\mu$eV as the unit of the $\Gamma$.


%
%


\subsection{back-action}
Usually, it is assumed that the energy-levels of QDs are not changed.
However, the energy-levels of QDs are changed in several situations.
For example, it can be considered that there are trap sites near the QDs, 
and the charge distribution of the trap site changes depending on the externally applied voltage.
In addition, we can consider the case of \cite{TanaAIP,Engel} where 
the energy-levels are affected by the directions of the spins that fill the lower energy-levels of the same QDs. 
In these cases, it is natural to consider that both $E_2$ and $E_4$ are changed by the measurements.
Thus, it is meaningful to analyze the effect of the measurements on those energy-levels.
Because the change of the energy-levels affects the electronic states of electrons,
it is related to the decoherence effect. 
In many literatures, the decoherence effects have been
analyzed regarding the noise effect on the coherence.
However, the detailed noise analysis of the qubits is complicated and requires 
a lot of experimental data~\cite{Tarucha}. 
Here, we consider that the decoherence in QD2 and QD4 is induced by the measurement
of the currents 1,3 and 5.
That is, 
it is possible that electrons in QD2 or QD4 lose their coherence while 
they move back and forth to the channel QDs 1,3,5. 
We simply describe the decoherence time caused by the interaction. 
This process can be described by the Golden rule~\cite{Schoelkopf},
where the last term of the Hamiltonian Eq.~(\ref{Hamilt}), 
$ H_{\rm int}(t)\equiv \sum_{i=1}^{4}\sum_s W_{i,i+1}(d_{i}^\dagger(t) d_{i+1}(t)+h.c.)$, 
is treated as the perturbation term.
Then, the relaxation rate $\Gamma_{\rm relax}$ can be defined by
\begin{eqnarray}
\Gamma_{\rm relax} &\approx &\frac{1}{\hbar^2} 
\int_{-\infty}^\infty d\tau e^{-i\omega_{01}\tau} 
\langle H_{\rm int}(\tau)H_{\rm int}(0) \rangle, 
\end{eqnarray}
where $\omega_{01}=|E_2-E_4|$.
The decoherence time $t_{\rm dec}$ is defined by $t_{\rm dec}\equiv \Gamma_{\rm relax}^{-1}$. 
The final form of $t_{\rm dec}$ is given by(see Appendix)
\begin{widetext}
\begin{eqnarray}
t_{\rm dec}^{-1}&\approx &
\sum_{i} \frac{ |W_{i,i+1}|^2 }{\hbar^2} 
 \int \frac{d\epsilon'}{2\pi} 
[ g_{di}^<(\epsilon' +\omega_{01})g_{di+1}^>(\epsilon')+ g_{di+1}^<(\epsilon' +\omega_{01})g_{di}^>(\epsilon')] 
\nnm \\
&=&
[1-f(E_2)]
\left( 
\frac{ |W_{12}|^2 }{\hbar^2}\frac{\Gamma_1^Lf_L(E_2+\omega_{01})+\Gamma_1^R f_R(E_2+\omega_{01})}
{[E_2+\omega_{01}-E_1-\Lambda_i(\omega)]^2+ \Gamma_1^2/4}  + (1\rightarrow 3) \right)\nnm \\
& &+ 
f(E_2) 
\left(
\frac{ |W_{12}|^2 }{\hbar^2}\frac{\Gamma_1^L(1-f_L(E_2-\omega_{01}))+\Gamma_1^R (1-f_R(E_2-\omega_{01}))}
{[E_2-\omega_{01}-E_1-\Lambda_i(\omega)]^2+ \Gamma_1^2/4}  + (1\rightarrow 3) \right)\nnm \\
& & +
[1-f(E_4)]
\left(
\frac{ |W_{43}|^2 }{\hbar^2}
\frac{\Gamma_1^Lf_L(E_4+\omega_{01})+\Gamma_1^R f_R(E_4+\omega_{01})}
{[E_4+\omega_{01}-E_3-\Lambda_i(\omega)]^2+ \Gamma_1^2/4}   + (3\rightarrow 5) \right)\nnm \\
& &+ 
f(E_4) 
\left(
\frac{ |W_{43}|^2 }{\hbar^2}\frac{\Gamma_1^L(1-f_L(E_4-\omega_{01}))+\Gamma_1^R (1-f_R(E_4-\omega_{01}))}
{[E_4-\omega_{01}-E_3-\Lambda_i(\omega)]^2+ \Gamma_1^2/4}   + (3\rightarrow 5) \right).
\label{tdec}
\end{eqnarray}
\end{widetext}
Here, $f(E_i)$ and $1-f(E_i)$ means that there is an electron in the $E_i$ level and 
that there is no electron in the $E_i$ level, respectively ($i=2,4$).
In order to treat an average case, we take $f(E_2)=f(E_4)=1/2$ in the following calculations.

\begin{figure}
\centering
\includegraphics[width=8.8cm]{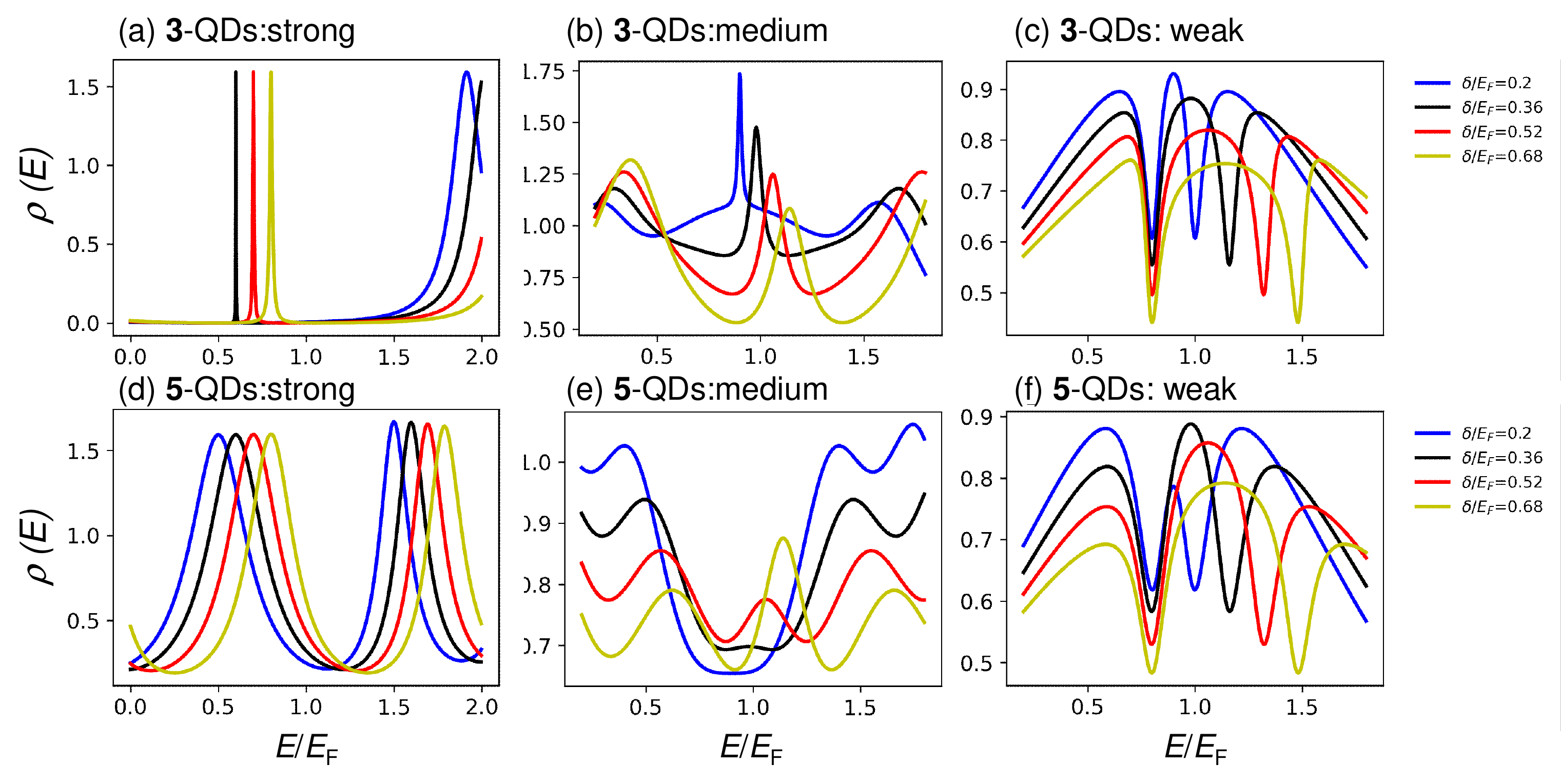}
\caption{The density of states (DOS), Eq.(\ref{eqDOS}), for 
(a)(d) the strong measurement case of $\Gamma=0.2\Gamma_0$ and $W=\Gamma_0$,
(b)(e) the medium measurement case of $\Gamma=W=0.5\Gamma_0$,
and (c)(f) the weak measurements case of $\Gamma=\Gamma_0$ and $W=0.2\Gamma_0$.
$\Gamma_0=10\mu$eV and $E_2/E_F=0.5$.
(a)(b) and (c) for the three-QD case. (d)(e) and (f) for the five-QD case.
}
\label{figDOS}
\end{figure}

\section{Numerical Results}\label{sec:results}
\begin{figure*}
\centering
\includegraphics[width=14.0cm]{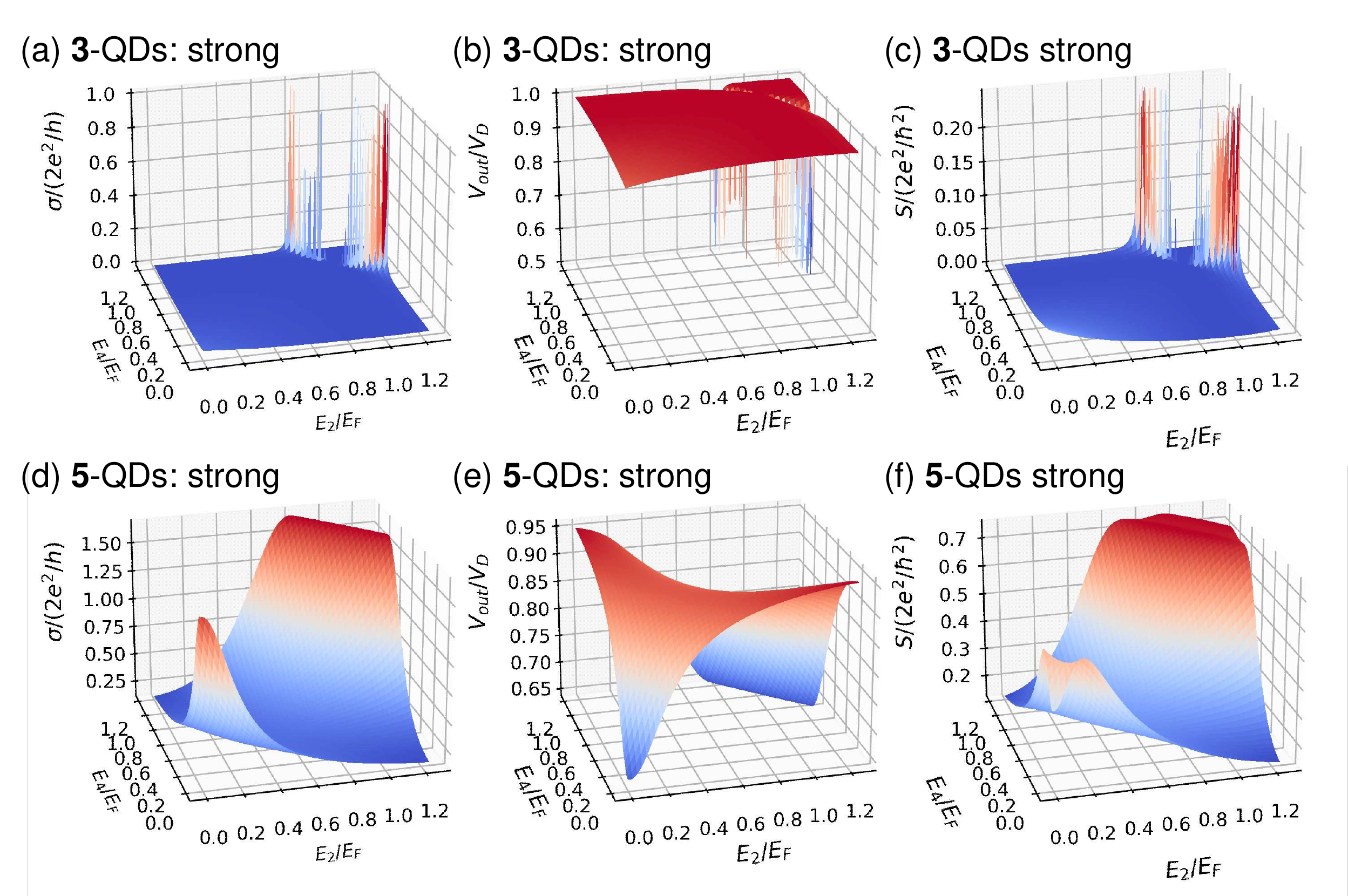}
\caption{The strong measurement case of $\Gamma=0.2\Gamma_0$ and $W=\Gamma_0$ 
as functions of $E_2$ and $E_4$ for several $\delta=(E_4-E_2)/2$.
$\Gamma_0=10\mu$eV and $V_{\rm D}=1$meV.
(a)(d) Conductance (b)(e) $V_{\rm out}$. (c)(f) The noise power $S$.
(a)(b)(c) Three-QD case and (d)(e)(f) five-QD case.}
\label{figGstrong}
\end{figure*}

\begin{figure*}
\centering
\includegraphics[width=14.0cm]{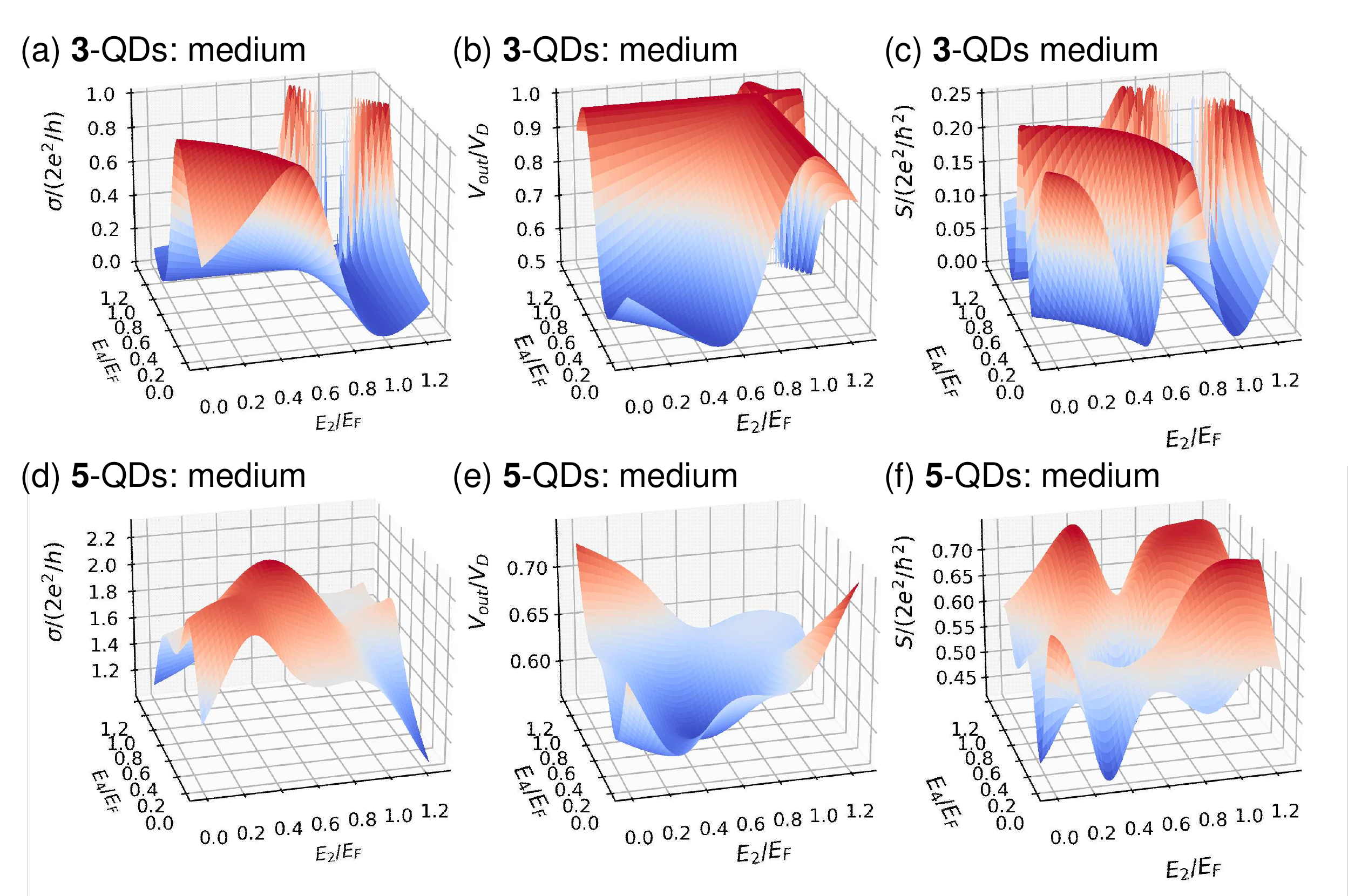}
\caption{The medium measurement case of $\Gamma=W=0.5\Gamma_0$
as functions of $E_2$ and $E_4$ for several $\delta=(E_4-E_2)/2$.
$\Gamma_0=10\mu$eV and $V_{\rm D}=1$meV.
(a)(d) Conductance (b)(e) $V_{\rm out}$. (c)(f) The noise power $S$.
(a)(b)(c) Three-QD case and (d)(e)(f) five-QD case.}
\label{figGmedium}
\end{figure*}

\begin{figure*}
\centering
\includegraphics[width=14.0cm]{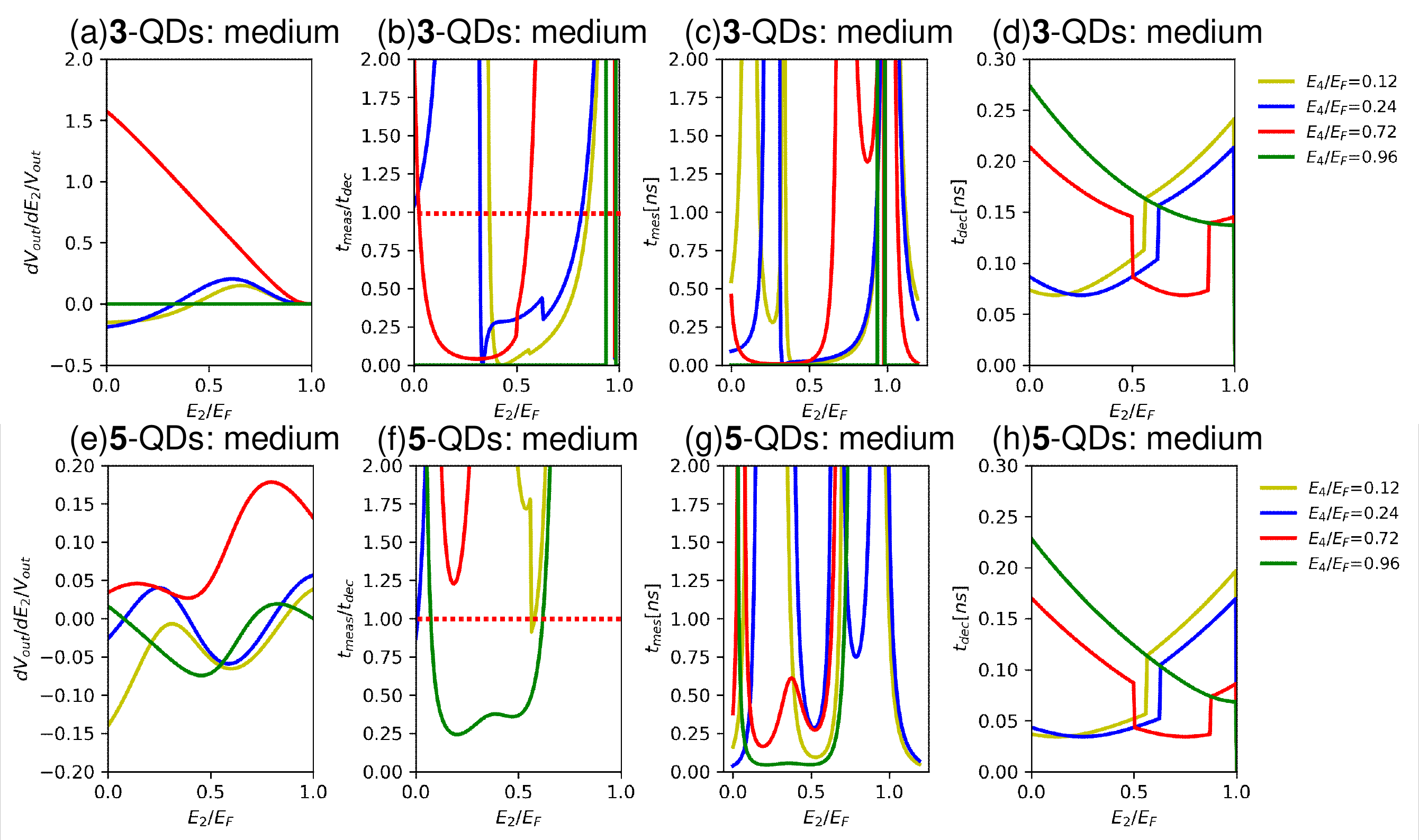}
\caption{The medium measurement case of $\Gamma=W=0.5\Gamma_0$
as functions of $E_2$ and $E_4$ for several $\delta=(E_4-E_2)/2$.
$\Gamma_0=10\mu$eV and $V_{\rm D}=1$meV.
(a)(e) $\frac{d V_{\rm out}}{d (E_2-E_4)}/V_{\rm D}$. 
(b)(f) $t_{\rm meas}/t_{\rm dec}$. 
(c)(g) $t_{\rm meas}$.
(d)(h) $t_{\rm dec}$.
(a)(b)(c)(d) Three-QD case and (e)(f)(g)(h) five-QD case.
The dotted lines in Figs.(b) and (f) indicate the boundary of the effective measurement $t_{\rm meas}<t_{\rm dec}$.
}
\label{figtmedium}
\end{figure*}

\begin{figure*}
\centering
\includegraphics[width=14.0cm]{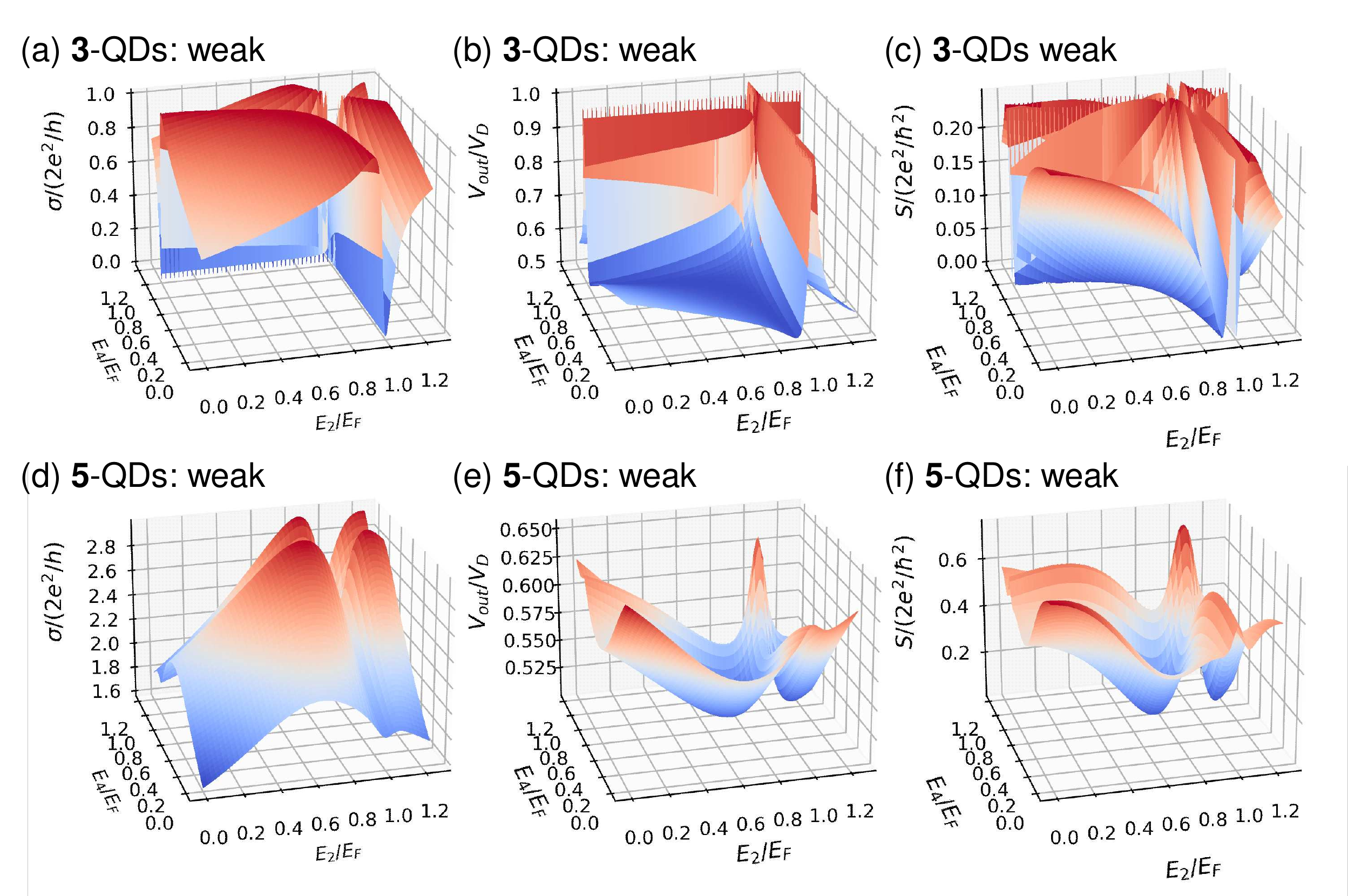}
\caption{The weak measurement case of $\Gamma=\Gamma_0$ and $W=0.2\Gamma_0$
as functions of $E_2$ and $E_4$ for several $\delta=(E_4-E_2)/2$.
$\Gamma_0=10\mu$eV and $V_{\rm D}=1$meV.
(a)(d) Conductance (b)(e) $V_{\rm out}$. (c)(f) The noise power $S$.
(a)(b)(c) Three-QD case and (d)(e)(f) five-QD case.}
\label{figGweak}
\end{figure*}

\begin{figure*}
\centering
\includegraphics[width=14cm]{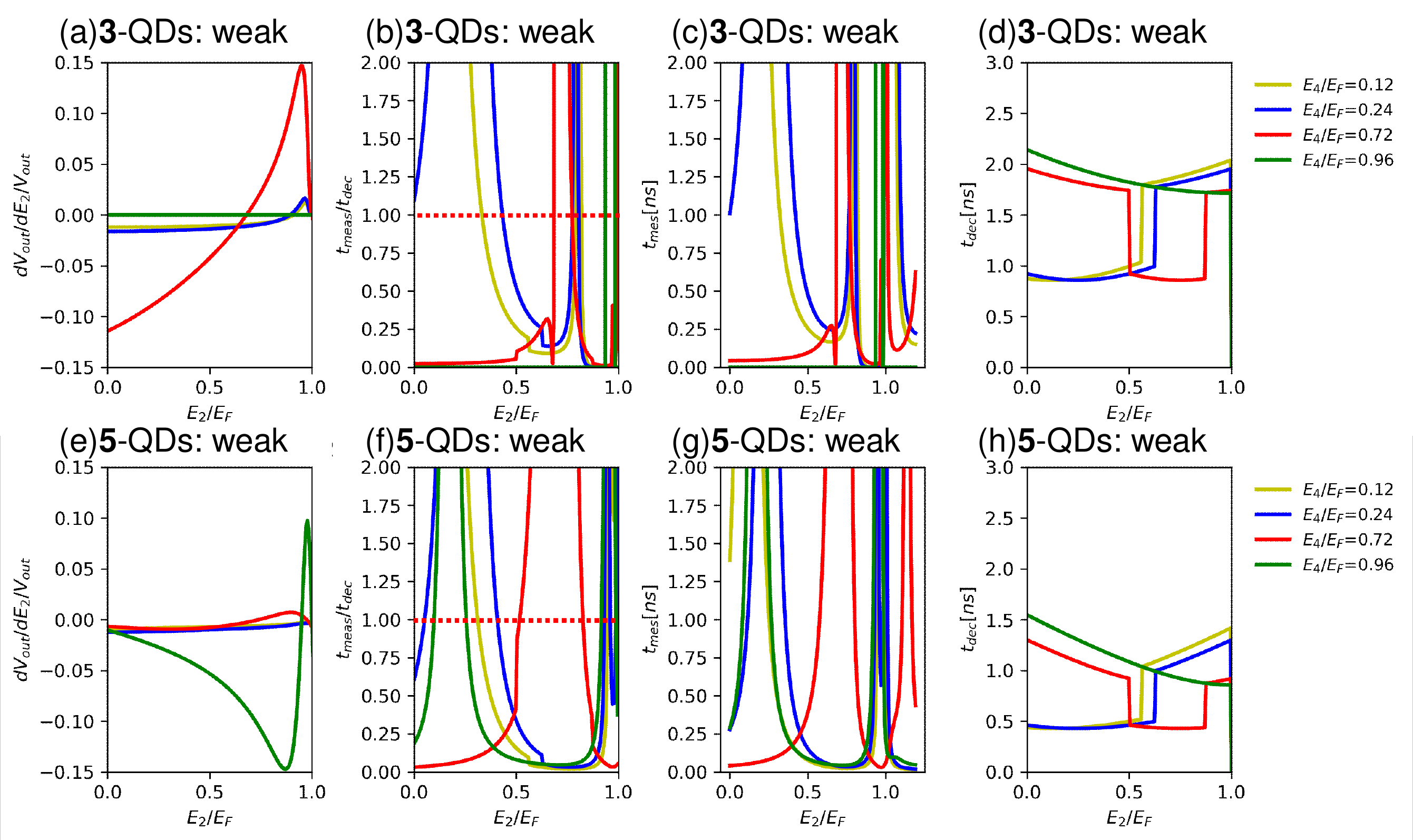}
\caption{The weak measurement case of $\Gamma=\Gamma_0$ and $W=0.2\Gamma_0$
as functions of $E_2$ and $E_4$ for several $\delta=(E_4-E_2)/2$.
$\Gamma_0=10\mu$eV and $V_{\rm D}=1$meV.
(a)(e) $\frac{d V_{\rm out}}{d (E_2-E_4)}/V_{\rm D}$. 
(b)(f) $t_{\rm meas}/t_{\rm dec}$. 
(c)(g) $t_{\rm meas}$.
(d)(h) $t_{\rm dec}$.
(a)(b)(c)(d) Three-QD case and (e)(f)(g)(h) five-QD case.
The dotted lines in Figs.(b) and (f) indicate the boundary of the effective measurement $t_{\rm meas}<t_{\rm dec}$.
}
\label{figtweak}
\end{figure*}

For simplicity, we assume the uniform case of $W_{ij}(=W)$ and $\Gamma_{i}(=\Gamma)$ at zero temperature ($T=0$).
When we use QDs 1,3, and 5, with their electrodes as the measurement structure 
to detect the energy-levels of QD 2 and 4,
the magnitude of $W$ compared with $\Gamma$ can be regarded as the strength of measurement.
Thus, we can distinguish the following three regions: \\
(1) {\it Strong} measurement of $W<\Gamma$ (Fig.~\ref{figGstrong}), \\
(2) {\it Medium} measurement of $W\approx\Gamma$ (Figs.~\ref{figGmedium} and~\ref{figtmedium}), \\
(3) {\it Weak} measurement of $W<\Gamma$ (Figs.~\ref{figGweak} and~\ref{figtweak}). \\

Since we focus on the detection of the difference of the two energy-levels,
the change of the currents from those at $E_2=E_4$ is important.
Thus, all numerical results are described as the functions of $E_2$ and $E_4$.
We also assume that the difference of the energy-levels between the adjacent QDs are uniform 
such that $\delta\equiv E_2-E_1=E_3-E_2=E_4-E_3=E_5-E_4$.

Figure~\ref{figDOS} shows the DOS of Eq.(\ref{eqDOS}) for the three coupling regions.
In the strong measurement case of Figs.~\ref{figDOS}(a)(d), the left peak shows the energy-level of the QD2
(we fix $E_2$ in the calculation), and the right peak shows the coupling to the electrodes.
In the medium measurement of Figs.~\ref{figDOS}(b)(e), 
we can see both the central sharp peak and the two broad peaks,
which are similar characteristics to those 
discussed in~\cite{Orellana1,Ulloa,Wang,Domanski,Orellana2}. 
In the weak coupling cases (Figs.~\ref{figDOS}(c)(f)), we observe the Fano dip structure over the broad Lorentzian structure.
When the three-QD medium coupling case (Fig.\ref{figDOS}(b)) is compared with that of five-QD case (Fig.\ref{figDOS}(e)), 
the peak structures are broadened.
This is because the five-QD structure has additional electrodes compared with the three-QD case,
and the coupling to the electrodes makes the Lorentzian wider.
In contrast, for the weak coupling case, 
there is no significant difference in both the three-QD case and the five-QD case.
This is because the coupling between the channel current and the electrodes are weak, 
resulting in the smaller effects of the additional electrodes of the five-QD structure. 
The effect of the increasing detuning $\delta=E_{i+1}-E_{i}$ is prominent 
in the case of the five-QD case for the medium measurement. 
This is because that there are two additional QDs in the case of the five-QD case in Fig.~\ref{figDOS}(e).

Figures~\ref{figGstrong} show the transport properties of the strong measurement case.
We can see that the conductances have the peak structures around the Fermi energy.
This can be understood by considering that the two kinds of peaks of Figs.~\ref{figDOS} (a) and (d)
overlap around the Fermi energy.  
The output $V_{\rm out}$ is in the same order of the applied voltage of $V_{\rm D}$.
Owing to the fact that the coupling $\Gamma$ to the electrodes is weaker than the coupling $W$ to the QDs,
the shot noises (Fig.~\ref{figGstrong}(c) and (f)) are smaller than those of the following 
medium and weak measurement cases.
A larger change in $V_{\rm out}$ as the function of the difference between $E_2$ and $E_4$ is desirable.
In this meaning, the strong measurement case shows the large $V_{\rm out}$.
However, this strong measurement case did not hold the condition
$t_{\rm meas} < t_{\rm dec}$, which implies that the measurement was completed during the decoherence time
(figures not shown).

Next, we consider the medium measurement case shown in Figs.~\ref{figGmedium} and \ref{figtmedium}.
In case of the three QDs,
it is observed that the conductance (Fig.~\ref{figGmedium}(a)) decreases before the Fermi level, 
and increases at the Fermi level.
This wall-like structure around $(E_2,E_4)=(0,0)$ of Fig.~\ref{figGmedium}(a) can be partly explained by 
considering the zero points of the denominator of
Eq.(\ref{J3QD}) given by 
\begin{eqnarray}
(1-C_{23}^r-C_{34}^r)D_3 
&=& b_3^*-W^2\left[\frac{b_2^* }{D_2}+\frac{b_4^*}{D_4}\right],
\label{wall} 
\end{eqnarray}
where $b_i(\omega)= \omega-E_i +i\Gamma_i/2$ 
and $D_i= [\omega-E_i]^2+\gamma_i^2$
($\gamma_i=\Gamma_i/2$). 
The zero points of Eq.(\ref{wall}) leads to
\begin{equation}
1 = \frac{W^2}{[\omega-E_2]^2 +\gamma^2}+\frac{W^2}{[\omega-E_4]^2 +\gamma^2}. 
\label{wall2}
\end{equation}
For $\Gamma=W$, Eq.(\ref{wall2}) satisfies $\cos^2 \theta_a+\sin^2 \theta_a=1$
if we take
\begin{eqnarray}
\omega-E_2 &=&\gamma \tan \theta_a, \\
\omega-E_4 &=&\gamma \cot \theta_a.
\end{eqnarray}
Thus, we obtain 
\begin{eqnarray}
(\omega-E_2)(\omega-E_4) &=&\gamma^2. 
\end{eqnarray}
This equation means that the maximum current is observed 
when $E_2$ and $E_4$ has the relation $y=1/x$ from the point $(x,y)=(E_F,E_F)$.
Compared with the three QD case, the results of the five QD case has 
a peak structure (Fig.~\ref{figGmedium}(d)).
This is because of the complicated structure of Eq.~(\ref{current5}).

\begin{figure*}
\centering
\includegraphics[width=14.0cm]{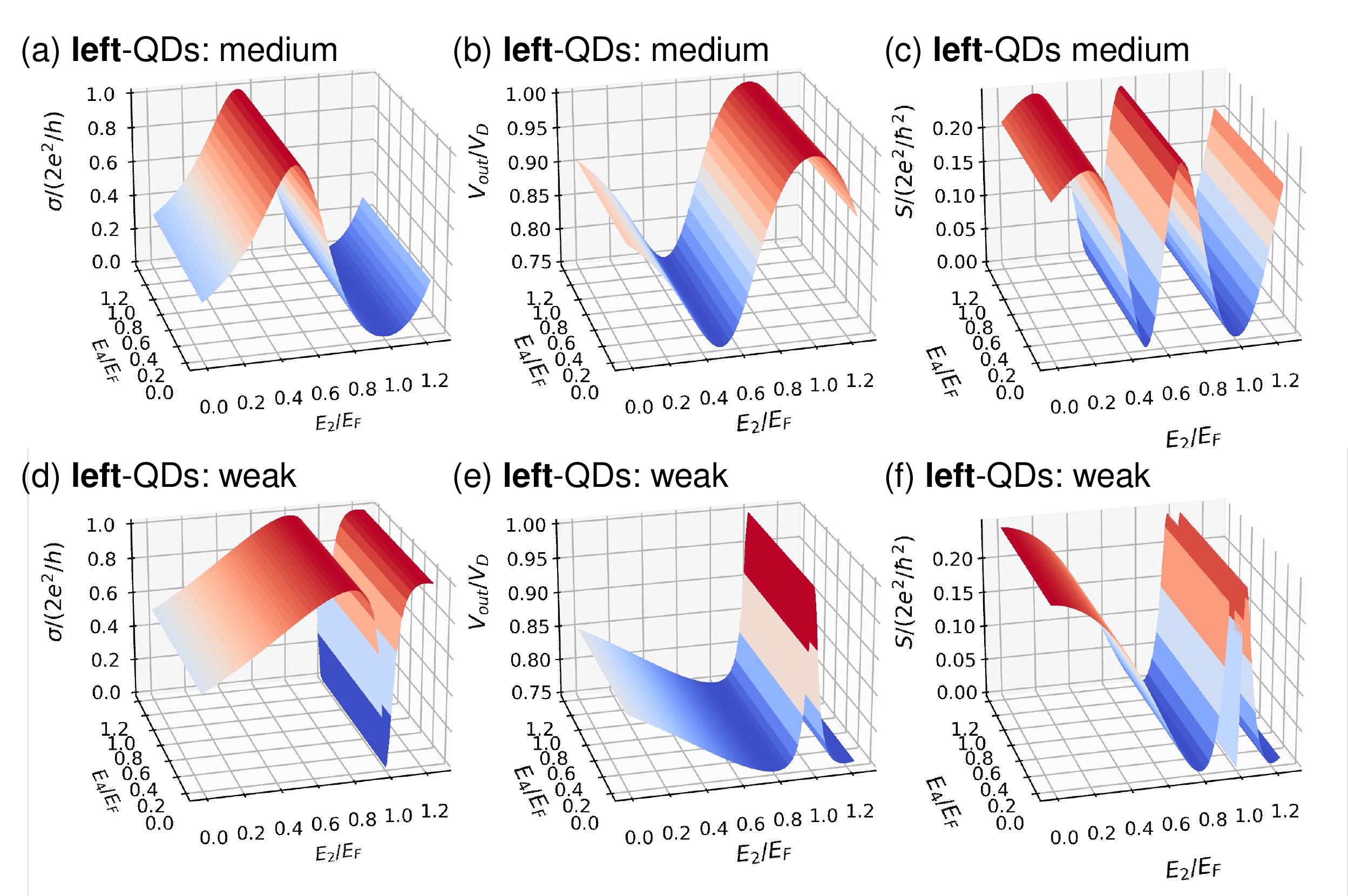}
\caption{The transport properties, where only the left current channel 1 is switched on.
(a)(b)(c) for the medium measurement case ($\Gamma=W=0.5\Gamma_0$), and
(d)(e)(f) for the weak measurement case ($\Gamma=\Gamma_0$ and $W=0.2\Gamma_0$).
(a)(d) Conductance (b)(e) $V_{\rm out}$. (c)(f) The noise power $S$.
$\Gamma_0=10\mu$eV and $V_{\rm D}=1$meV.}
\label{figleftcurrent}
\end{figure*}

Figure~\ref{figGmedium} (b) shows that 
the large change of $V_{\rm out}$ can be seen around the wall-like structure for the three-QD case, 
and Figure~\ref{figGmedium} (e) shows that 
$V_{\rm out}$ changes prominently away from the diagonal line of $E_2=E_4$.
Accordingly, Figures~\ref{figtmedium} (a) and (e) shows the large rate of $\frac{d V_{\rm out}}{d (E_2-E_4)}/V_{\rm D}$ 
around the middle of the Fermi surface ($E_4/E_F=0.72$). 
Compared with Figs.~\ref{figtmedium}(a) and (e), 
the three-QDs have a higher amplifying rate.
Because $V_D=1$ meV in the present case, we can expect 
$V_{\rm out}$ changes more than 1 meV when $E_2$ changes for a fixed $E_4$
for the three-QD case.
A comparison of Fig~\ref{figGmedium}(c) and Fig~\ref{figGmedium}(f) 
shows that the coupling of the five-QD case induces a larger noise than the three-QD case.
This is because of the three current lines attached to the QDs for the five-QD case.
From Fig.~\ref{figtmedium}(d) and (h), 
the decoherence time of the three-QD case is a little longer than that of the five-QD case.
Here, the abrupt change of $t_{\rm dec}$ comes from the definition of $|\omega_{01}|$ in
Eq.(\ref{tdec}) (see Appendix and Fig.~\ref{figtabrupt}).
When the measurement times are compared 
(Figs.~\ref{figtmedium} (c) and (g)),
the measurement time $t_{\rm meas}$ of the five-QD case is 
longer than that of the three-QD case.
Here, in the calculation of $t_{\rm meas}$, Eq.(\ref{eqmeasurement}), 
we exclude the line around $E_2=E_4$, which results in the divergence structures in Figs.~\ref{figtmedium} (b)(c)(f)(g).
The dot lines in Figs.~\ref{figtmedium}(b) and (f) show the boundary line of the effective measurement 
mentioned above. 
That is, the setup of $t_{\rm meas} \ge t_{\rm dec}$ is meaningless because 
before obtaining the information of the energy-levels of QD2 and QD4, 
the electrons dephase via the back-action of the measurement.
The three-QD case satisfies $t_{\rm meas} \ll t_{\rm dec}$, 
and the measurement is effective for most of the $E_2$ region.
Thus, the side-QD setup of Fig.~1(a) is a good measurement apparatus for the energy-levels of the two QDs, 
whereas the five-QD case is meaningless in most of the region of $E_2$.

As seen in Fig.~\ref{figDOS}, the weak measurement case shows the Fano dip structure, 
and the conductances (Figs.~\ref{figGweak}(a) and (d)) reflect the corresponding dip structures around $E_F$. 
$V_{\rm out}$ shows sharp changes around $E_F$ in both Figs.~\ref{figGweak}(b) and (e).
From Figs.~\ref{figtweak}(a) and (e), the five-QD case shows similar amplifying performances to the three-QD case. 
The decoherence time of the three-QD case is better than that of the five-QD case, 
and accordingly, the measurement time $t_{\rm meas}$ and the ratio $t_{\rm meas}/t_{\rm dec}$
of the three-QD case is a little better than those of the five-QD case. 
From the measurement and back-action viewpoint, it seems that the three-QD case is a little better than the five-QD case
for the weak measurement. 

Figures~\ref{figleftcurrent} show the case where only the current channel 1 of Fig.1(b) is switched on.
In this case, the current reflects only the energy-level of the QD2.
The conductance behaves differently from the numerical results mentioned above, 
and by comparing the current where only current 3 is switched on, 
we could distinguish which of the energy-levels between QD2 and that of QD4 were larger.

\section{Discussion}\label{sec:discussion}
As an application of the present structure, 
we think that our setup can be used 
to distinguish the two Zeeman energies.
Let us estimate the concrete values of the energy-difference of the Zeeman energies 
shown in~\cite{Takeda} and \cite{Struck}.
The gradient magnetic field in~\cite{Takeda} described the 30 mT magnetic field gradient between two QDs approximately 100 nm apart, 
and 0.08 mT/nm in~\cite{Struck}, which corresponds to the 8 mT between the two QDs 100 nm apart.
When we estimate the Zeeman energy for the magnetic field gradient $\Delta B_z$ [T] by 
$\Delta E_Z=g\mu_B \Delta B_z$ with $g=2$ and $\mu_B=5.789\times 10^{-4}$ [eV/T],
the energy-difference of $\Delta B_z$=30mT corresponds to $\Delta E_Z=3.47$ meV.
For the Fermi energy $E_F=\hbar^2 k_F^2/(2m)$ with $k_F=(3\pi^2 n_e)^{1/3}$ 
($m$ is the effective mass and $n_e$ is the charge density), 
when we take $m=0.5 m_0$ (Si) and $n_e=10^{19}$ cm$^{-3}$, 
we have $E_F=15.7$ meV. 
When we take $m=0.067m_0$ (GaAs), we have $E_F=1.17$eV.
Thus $E_2-E_4=\Delta E_Z/E_F \approx 0.22$ and 0.03, respectively, and 
they are in the range of the present numerical calculations (here we chose $\Gamma_0=10\mu$eV and $V_D$=1 meV).
In the present medium measurement case (Figs.~\ref{figtmedium}(a) and (e)), 
the change of $V_{\rm out}$ is in order of the 0.01 meV$\sim$ 1 meV and detectable~\cite{TanaAPL}.
Thus, the medium measurement region is good for detecting energy-levels in the present parameter setting.
The numerical results changes depending on the parameter regions,
and if we choose the parameters appropriately, we will be able to detect the energy-difference. 
In order to directly compare the numerical results with experiments, we need to adjust the parameters.

\section{Conclusion}\label{sec:conclusion}
We have theoretically investigated the three and five QD systems as the measurement system 
of energy-levels of internal QDs by 
considering the additional small circuit to convert the 
current changes into voltage changes.
We observed that depending on the coupling strength of 
the measurement part and the targeted internal QDs, 
the conductance, noise, and output voltage changes.
We have also estimated the measurement time and the decoherence time,
and showed the trade-off between the measurement strength and the 
decoherence time.
It was found that the medium measurement region is good for the 
detection of the difference between two energy-levels.
It was also found that the three-QD case shows a wide range of effective measurement 
regions compared with the five-QD case. 

\begin{acknowledgements}
We are grateful to K. Ono, T. Mori and H. Fuketa for their fruitful discussions.
This work was partly supported by MEXT Quantum Leap Flagship Program (MEXT Q-LEAP) Grant Number JPMXS0118069228, Japan.
The authors thank the Supercomputer Center, the Institute for
Solid State Physics, and the University of Tokyo for the use of their facilities.
\end{acknowledgements}

\appendix

\section{Dephasing rate}
\begin{figure}
\centering
\includegraphics[width=8cm]{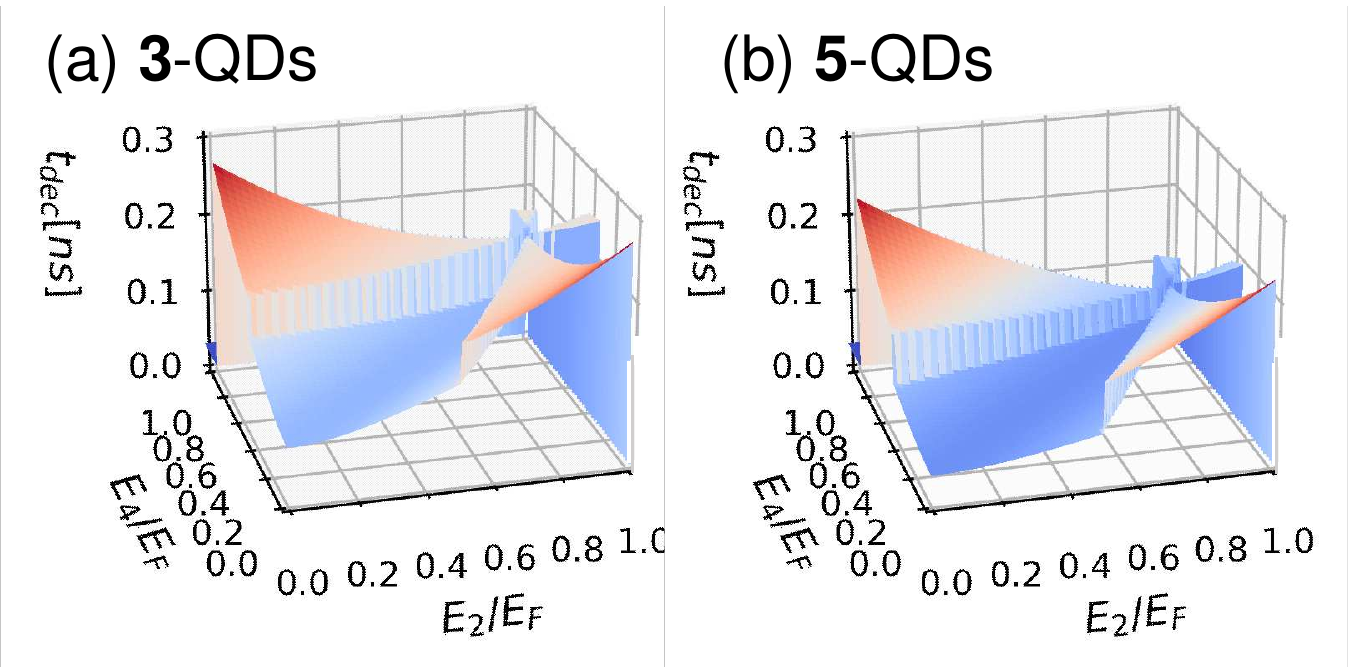}
\caption{3D view of the decoherence time for the medium measurement case of $\Gamma=W=0.5\Gamma_0$ for $\Gamma_0=10\mu$eV and $V_{\rm D}=1$meV.
(a)The three-QD case and (b) the five-QD case.
The abrupt changes of Figs.~\ref{figtmedium} and \ref{figtweak} comes from eqs.(\ref{abrupt1}) and (\ref{abrupt2}). 
}
\label{figtabrupt}
\end{figure}

The dephasing rate described by the Golden rule~\cite{Schoelkopf} 
can be calculated from the correlation function $\langle H_{\rm int}(t)H_{\rm int}(0) \rangle$,
which is given by
\begin{eqnarray}
\!\! \! \lefteqn{ \langle H_{\rm int}(t)H_{\rm int}(0) \rangle \!\!=\!\!
\langle \sum_{ij}\!(W_{i,i+1}d_i^\dagger (t\!) d_{i+1}(t\!)\!\!+\! W_{i,i+1}^* d_{i+1}^\dagger (\! t) d_i(\! t)) }\nnm\\
&\times&(W_{j,j+1} d_j^\dagger (0) d_{j+1}(0)+W_{j,j+1}^* d_{j+1}^\dagger (0) d_j(0))
\rangle \nnm \\
\!\!&=&
\sum_{i} |W_{i,i+1}|^2 
\langle d_i^\dagger (t) d_i(0)\rangle \langle d_{i+1}(t)d_{i+1}^\dagger (0) \rangle
\nnm \\
&+& |W_{i,i+1}|^2 
 \langle d_{i+1}^\dagger (t) d_{i+1}(0)\rangle \langle d_i (t) d_{i}^\dagger(0)\rangle
\nnm \\
\!\!&=&\!\!
\sum_{i} |W_{i,i+1}|^2 
[g_{di}^<(-t) g_{di+1}^> (t) + g_{di+1}^<(-t)g_{di}^> (t)]
 \nnm \\
\!\!\!\!&=&\!\!\!\!
\sum_{i} |W_{i,i+1}|^2  \!\!\!\!
\int\!\!\!\!\int\!\!  \frac{d\epsilon}{2\pi}\frac{d\epsilon'}{2\pi} 
[ g_{di}^<(\epsilon)g_{di+1}^>(\epsilon')\!+\! g_{di+1}^<(\epsilon)g_{di}^>(\epsilon')] 
 e^{i(\epsilon -\epsilon' )t}. 
 \nnm \\ \!\!\!
\end{eqnarray}
Then, the relaxation rate is given by
\begin{eqnarray}
\lefteqn{
\Gamma_{\rm relax} \approx 
\frac{1}{\hbar^2} \int_{-\infty}^\infty d\tau e^{-i\omega_{01}\tau} 
\langle H_{\rm int}(\tau)H_{\rm int}(0) \rangle} \nnm \\
&=& \sum_{i} \frac{ |W_{i,i+1}|^2 }{\hbar^2} 
 \!\int\!\! \frac{d\epsilon'}{2\pi}\! 
[ g_{di}^<(\epsilon' +\omega_{01})g_{di+1}^>(\epsilon') \nnm \\
&+& g_{di+1}^<(\epsilon' +\omega_{01})g_{di}^>(\epsilon')]. 
\end{eqnarray}
The abrupt change $t_{\rm dec}$ originates from the definition of $\omega_{01}$. 
For $E_2>E_4$, $\omega_{01}=E_2-E_4$, and we have
\begin{eqnarray}
E_2+\omega_{01} &=& 2E_2-E_4, \ E_2-\omega_{01} = E_4, \\
E_4+\omega_{01} &=& E_2, \ E_4-\omega_{01} = 2E_4-E_2. 
\label{abrupt1} 
\end{eqnarray}
For $E_2<E_4$, $\omega_{01}=E_4-E_2$, and we have
\begin{eqnarray}
E_2+\omega_{01} &=& E_4, \ E_2-\omega_{01} = 2E_2-E_4, \\
E_4+\omega_{01} &=& 2E_4-E_2, \ E_4-\omega_{01} = E_2. 
\label{abrupt2}
\end{eqnarray}
The bird-eye views of $t_{\rm dec}$ are shown in Fig.~\ref{figtabrupt}.

\section{Green functions for the QDs}
In this section, we show the derivation of the Green functions 
based on the equation of motion method.
From Eq.(\ref{eqmotion}), we have the equations like
\begin{equation}
(\omega-E_{1}) d_{1} 
= \sum_{k_L} V_{k_{L}1}^* c_{1k_{L}}+\sum_{k_R}V_{k_{R}1}^* c_{1k_{R}} +W_{12}d_{2}.
\end{equation}

Thus, we obtain
\begin{eqnarray}
\!& &\!\left(\omega\!-\! E_{1}\!-\!\Sigma_1(\omega)\right) G_{d_{1},d_{j}}
= W_{12}G_{d_{2},d_{j}}+\delta_{1j}, \\
\!& &\! \left(\omega\!-\! E_{2} \right) G_{d_{2},d_{j}}
= W_{21}G_{d_{1},d_{j}} +W_{23}G_{d_{3},d_{j}}+\delta_{2,j}, \\
\!& &\! \left(\omega\!-\! E_{3}\!-\Sigma_3 (\omega) \right) G_{d_{3},d_{j}}\!
 \!=\! W_{32}G_{d_{2},d_{j}}\!\!+\!W_{34}G_{d_{4},d_{j}}\!\!+\!\delta_{3,j}, \nnm\\ \\
\!& &\! \left(\omega\!-\! E_{4} \right) G_{d_{4},d_{j}}= W_{45}G_{d_{5},d_{j}} +W_{43}G_{d_{3},d_{j}}+\delta_{4,j}, \\
\!& &\!\left(\omega\!-\! E_{5}\!-\!\Sigma_5(\omega)\right) G_{d_{5},d_{j}}= W_{54}G_{d_{4},d_{j}}+\delta_{4j},
\end{eqnarray}
where
\begin{equation}
\Sigma_i(\omega) \equiv  \sum_{k_L} \frac{|V_{k_{Li}}|^2}{(\omega-E_{k_{Li}})}
         +\sum_{k_R}\frac{|V_{k_{Ri}}|^2}{(\omega-E_{k_{Ri}})}.
\end{equation}
Thus, the equations between the Green functions of the QDs are given as follows:
\begin{eqnarray}
G_{d11} &=& W_{12}g_{d1}G_{d21}+g_{d1}, \\
G_{d21} &=& C_{12}G_{d21} +W_{21}g_{d2}g_{d1} +W_{23}g_{d2} G_{d31}, \\
G_{d22} &=& C_{12}G_{d22} +W_{23}g_{d2} G_{d32}+g_{d2},  \\
G_{d23} &=& C_{12}G_{d23} +W_{23}g_{d2} G_{d33},  \\
G_{d24} &=& C_{12}G_{d24} +W_{23}g_{d2} G_{d34},  \\
G_{d25} &=& C_{12}G_{d25} +W_{23}g_{d2} G_{d35},  \\
G_{d31} &=& g_{d3} (W_{32}G_{d21}+W_{34}G_{d41}), \\
G_{d32} &=& g_{d3} (W_{32}G_{d22}+W_{34}G_{d42}), \\
G_{d33} &=& g_{d3} (W_{32}G_{d23}+W_{34}G_{d43}+1), \\
G_{d34} &=& g_{d3} (W_{32}G_{d24}+W_{34}G_{d44}), \\
G_{d35} &=& g_{d3} (W_{32}G_{d25}+W_{34}G_{d45}), \\
G_{d41}&=& C_{54}G_{d41}+W_{43}g_{d4}G_{d31},  \\
G_{d42}&=& C_{54}G_{d42}+W_{43}g_{d4}G_{d32},  \\
G_{d43}&=& C_{54}G_{d43}+W_{43}g_{d4}G_{d33},  \\
G_{d44}&=& C_{54}G_{d44}+W_{43}g_{d4}G_{d34} +g_{d4}, \\
G_{d45}&=& C_{54}G_{d45}+W_{45}g_{d4}g_{d5}+W_{43}g_{d4}G_{d35},  \\
G_{d55}&=& W_{54}g_{d5}G_{d45}+g_{d5},  
\end{eqnarray}
where
\begin{eqnarray}
g_{di}(\omega) &=&1/(\omega- E_i -\Sigma_i).
\end{eqnarray}
These equations can be solved by starting with the elimination of
$G_{d2i}$ and $G_{d4i}$ ($i=1,...,5$) such as 
\begin{eqnarray}
G_{d21} &=& \frac{W_{21}g_{d2}g_{d1}}{(1-C_{12})} +\frac{W_{23}g_{d2} G_{d31}}{(1-C_{12})}, \\
G_{d22} &=& \frac{W_{23}g_{d2} G_{d32}}{(1-C_{12})}+\frac{g_{d2}}{(1-C_{12})}, \\
 & & .... \nnm
\end{eqnarray}
Thus, we have
\begin{eqnarray}
G_{d31} &=& \frac{(1-C_{54})}{\Delta_c}W_{32}W_{21}g_{d3}g_{d2}g_{d1}, \\
G_{d32} &=& \frac{(1-C_{54})}{\Delta_c}W_{32}g_{d3}g_{d2},\\
G_{d33} &=& \frac{(1-C_{12})(1-C_{54})}{\Delta_c}g_{d3}, \\
G_{d34} &=& \frac{(1-C_{12})}{\Delta_c}W_{34}g_{d3}g_{d4},\\
G_{d35} &=& \frac{(1-C_{12})}{\Delta_c}W_{34}W_{45}g_{d3}g_{d4}g_{d5}.
\end{eqnarray}
Here,
\begin{equation}
\Delta_c \equiv 
(1-C_{12}-C_{32})(1-C_{54})-(1-C_{12})C_{34}. 
\end{equation}
Similarly, we have
\begin{eqnarray}
G_{d11}&=& \frac{(1-C_{32})(1-C_{54})-C_{34}}{\Delta_c} g_{d1}, \\
G_{d55}&=& \frac{(1-C_{12})(1-C_{34})-C_{32}}{\Delta_c}g_{d5}. 
\end{eqnarray}
There are the Green functions of Eqs.(\ref{ggd11})-(\ref{ggd55}) in the main text.

\section{Green function for the QD-lead ($k$-$d$) elements}
Similar to the Green functions of the QDs, 
we can derive the Green functions of the type of $G_{d_i,k_{\alpha j}}(\omega) $
($ \equiv G_{d_i,c_{jk_{\alpha}}}(\omega)$ in Eq.(\ref{Gfundk}) in the main text) 
based on the equation of motion method as follows:
\begin{eqnarray}
& &\left(\omega-E_{1}-\Sigma_1(\omega)\right)G_{d_{1},k'_{\alpha j}}=W_{12}G_{d_{2},k'_{\alpha j}}
 +v_{k'_{\alpha 1}}\delta_{1j}, \\
\!& &\!\left(\omega-E_{2} \right) G_{d_{2},k'_{\alpha j}}\!=\! W_{21}G_{d_{1},k'_{\alpha j}}\!+\!W_{23}G_{d_{3},k'_{\alpha j}},\\
\!& &\!\left(\omega-E_{3}-\Sigma_3 (\omega)\right)G_{d_{3},k'_{\alpha j}}
= W_{32}G_{d_{2},k'_{\alpha j}}
+W_{34}G_{d_{4},k'_{\alpha j}} \nnm\\
& & \ \ +v_{k'_{\alpha 3}}\delta_{3j},\\
\!& &\!\left(\omega-E_{4} \right) G_{d_{4},k'_{\alpha j}}=W_{45}G_{d_{5},k'_{\alpha j}}+W_{43}G_{d_{3},k'_{\alpha j}},\\
\!& &\!\left(\omega-E_{5}-\Sigma_5(\omega)\right) G_{d_{5},k'_{\alpha j}}\!=\! W_{54}G_{d_{4},k'_{\alpha j}}\!+\!v_{k'_{L5}}\delta_{5j},
\end{eqnarray}
where ($j=1,3,5)$, and
\begin{eqnarray}
v_{k'_{\alpha i}}(\omega)&\equiv& \frac{V_{k'_{\alpha i}}^* }{(\omega-E_{k'_{\alpha i}})}. 
\end{eqnarray}
Hereafter we write the QD-lead Green functions $G_{d_{i},k_{\alpha j}}$ as $G_{ij}$, 
and $v_i \equiv v_{k'_{\alpha i}}(\omega)$ for simplicity.
These equations are changed into the following:
\begin{eqnarray}
G_{1j} &=& g_{d1}(W_{12}G_{2j} +v_1\delta_{ij}), \label{dk1}\\
G_{2j} &=& g_{d2}(W_{21}G_{1j}+W_{23}G_{3j}), \label{dk2}\\
G_{3j} &=& g_{d3}(W_{32}G_{2j}+W_{34}G_{4j} +v_3\delta_{3j} ),\label{dk3} \\
G_{4j} &=& g_{d4}(W_{45}G_{5j}+W_{43}G_{3j} ), \label{dk4}\\
G_{5j} &=& g_{d5}(W_{54}G_{4j}  +v_5). \label{dk5}
\end{eqnarray}
For the three Green functions $A(t,t')$,$B(t,t')$ and $C(t,t')$, 
if $A^r(t,t')=\int dt_1 B^r(t,t_1)C^r(t_1,t')$ is held,
we have the relation
\begin{eqnarray}
A^<(t,t')&=&\int dt_1 [B^r(t,t_1)C^<(t_1,t') \nnm \\
 & &+B^<(t,t_1)C^a(t_1,t')].
\end{eqnarray} 
When we apply this relation to Eq.(\ref{dk3}), we have
\begin{eqnarray}
&\! &G_{13}^<\!-\!C_{12}^r G_{13}^<=\! W_{23}W_{12} c_{12}^r G_{33}^<\! +\! C_{12}^< G_{13}^a\! +\! W_{23}W_{12} c_{12}^< G_{33}^a,\nnm \\ \\
&\! &G_{33}^<\! -[C_{32}^r+C_{34}^r]G_{33}^< = 
 W_{21}W_{32}c_{32}^r G_{13}^<+W_{21}W_{32}c_{32}^< G_{13}^a \nnm\\
&\! &+ \!W_{45}W_{34}c_{34}^r G_{53}^<\!+\!W_{45}W_{34}c_{34}^< G_{53}^a 
 \!+\![C_{32}^<+C_{34}^<]G_{33}^a
\! +[g_{d3}v_3]^<, \nnm \\ \\
&\! &G_{53}^<\!-\! C_{54}^r G_{53}^< = W_{43}W_{54}c_{54}^r G_{33}^<\! +\! C_{54}^< G_{53}^a
\!+\!W_{43}W_{54}c_{54}^< G_{33}^a, \nnm \\ 
\end{eqnarray}
where 
\begin{eqnarray}
c_{ij}^<&=& (g_{di}g_{dj})^<=g_{di}^r g_{dj}^<+g_{di}^< g_{dj}^a, \\
C_{ij}^<&=&|W_{ij}|^2 c_{ij}^<.
\end{eqnarray}
Here, we use the expressions of Eqs.(\ref{ggd11})-(\ref{ggd55}).
For example,
\begin{eqnarray}
C_{12}^< G_{13}^a + W_{23}W_{12} c_{12}^< G_{33}^a 
&=& \frac{(1-C_{54}^a)}{\Delta_c^a} W_{23}W_{12} c_{12}^< g_{d3}^av_3^a, \nnm \\ \\
W_{21}W_{32}c_{32}^< G_{13}^a + C_{32}^< G_{33}^a  
&=& \frac{(1-C_{54}^a)}{\Delta_c^a}  C_{32}^<  g_{d3}^av_3^a, \\
C_{54}^< G_{53}^a +W_{43}W_{54}c_{54}^< G_{33}^a
&=&
\frac{(1-C_{12}^a)}{\Delta_c^a}  C_{34}^<  g_{d3}^av_3^a. 
\end{eqnarray}
Then, we have
\begin{eqnarray}
 G_{13}^< &=& W_{23}W_{12} \frac{ c_{12}^r G_{33}^< + c_{12}^< g_{133}^a}{(1-C_{12}^r)}, \\
 G_{53}^< &=& W_{43}W_{54} \frac{ c_{54}^r G_{33}^< + c_{54}^< g_{533}^a}{(1-C_{54}^r)},
\end{eqnarray}
where 
\begin{eqnarray}
g_{133}^a
&=& \frac{(1-C_{54}^a)}{\Delta_c^a} g_{d3}^av_3^a, \\
g_{533}^a  &=& \frac{(1-C_{12}^a)}{\Delta_c^a} g_{d3}^a.
\end{eqnarray}

\begin{widetext}

Thus, we have
\begin{eqnarray}
G_{33}^<
&=& 
 |1-C_{54}^r|^2 \frac{[C_{32}^r C_{12}^< + C_{32}^<(1-C_{12}^r)]g_{d3}^av_3^a -(1-C_{12}^r)C_{32}^a[g_{d3}v_3]^<}{|\Delta_c^r|^2}  
\nnm \\
&+&
 |1-C_{12}^r|^2 \frac{[C_{34}^r C_{54}^< + C_{34}^<(1-C_{54}^r)]g_{d3}^av_3^a- (1-C_{54}^r)C_{34}^a[g_{d3}v_3]^<}{|\Delta_c^r|^2}  
+|1-C_{54}^r|^2|1-C_{12}^r|^2 \frac{[g_{d3}v_3]^<}{|\Delta_c^r|^2} 
\nnm \\
&=& 
 |1-C_{54}^r|^2 
 \frac{|W_{23}|^2|W_{12}|^2|g_2|^2 |g_3|^2 (g_1^< v_3^a + g_1^r v_3^<)
+|W_{32}|^2 |g_3^r|^2 [ g_2^< v_3^a -g_2^a v_3^<] }{|\Delta_c^r|^2}  
\nnm \\
&+&
 |1-C_{12}^r|^2 \frac{|W_{43}|^2|W_{54}|^2|g_4|^2 |g_3|^2 (g_5^< v_3^a + g_5^r v_3^<)
+|W_{34}|^2 |g_3^r|^2 [ g_4^< v_3^a -g_4^a v_3^<] }{|\Delta_c^r|^2}  
\nnm \\
&+& |1-C_{54}^r|^2|1-C_{12}^r|^2 \frac{[g_{3}^r v_3^< + g_{3}^< v_3^a]}{|\Delta_c^r|^2}.
\label{G33d}
\end{eqnarray}
\end{widetext}

Next, we consider the derivation of $G_{55}$ from Eq.(\ref{dk5}):
\begin{eqnarray}
& & G_{15}^<\!-\! C_{12}^r G_{15}^< = C_{12}^< G_{15}^a \!+\!W_{23}W_{12}c_{12}^r G_{35}^<\! +\! W_{23}W_{12}c_{12}^< G_{35}^a, \nnm\\
\label{g55_1}\\
& & G_{35}^< -[C_{32}^r+C_{34}^r]G_{35}^< = [C_{32}^<+C_{34}^<]G_{35}^a+W_{21}W_{32} c_{32}^r G_{15}^< \nnm \\
& &+W_{21}W_{32} c_{32}^< G_{15}^a
+W_{45}W_{34} c_{34}^r G_{55}^<+W_{45}W_{34} c_{34}^< G_{55}^a, \label{g55_2}\\
& & G_{55}^<-C_{54}^r G_{55}^< =C_{54}^< G_{55}^a 
+ W_{43}W_{54}c_{54}^r G_{35}^< \nnm\\
& &+ W_{43}W_{54}c_{54}^< G_{35}^a    +(g_{d5}v_5)^<.\label{g55_3}
\end{eqnarray}
Eq.(\ref{g55_1}) is changed to
\begin{equation}
 G_{15}^< = \frac{W_{23}W_{12}c_{12}^r}{(1-C_{12}^r)} G_{35}^< +\frac{C_{12}^<}{(1-C_{12}^r)}  g_{135}^a, 
\end{equation}
where 
\begin{eqnarray}
g_{135}^a &\equiv&\frac{W_{23}W_{12}W_{45}W_{34}}{|W_{12}|^2}\frac{c_{34}^a }{\Delta_c^a} (g_{d5}v_5)^a.
\end{eqnarray}
Eq.(\ref{g55_2}) is changed to
\begin{eqnarray}
& &\frac{(1-C_{34}^r)(1-C_{12}^r) -C_{32}^r}{(1-C_{12}^r)} G_{35}^<
= W_{45}W_{34} c_{34}^r G_{55}^< \nnm \\
& &+[C_{32}^< +\frac{C_{32}^r C_{12}^<}{(1-C_{12}^r)}  ] h_{135}^a 
+C_{34}^<g_{535}^a,   \nnm 
\end{eqnarray}
where
\begin{equation}
g_{535}^a \equiv   \frac{W_{45}W_{34} [ 1-C_{12}^a-C_{32}^a]}{|W_{34}|^2\Delta_c^a} g_{d5}v_5^a.
\end{equation}

Eq.(\ref{g55_3}) is changed to
\begin{equation}
(1 -C_{54}^r) G_{55}^< 
=W_{43}W_{54}c_{54}^r G_{35}^< +C_{54}^< h_{535}^a +(g_{d5}v_5)^<,
\end{equation}
where 
\begin{eqnarray}
h_{535}^a &\equiv & 
\frac{1-C_{12}^a-C_{32}^a }{\Delta_c^a} (g_{d5}v_5)^a.
\end{eqnarray}
Thus, we have
\begin{eqnarray}
\lefteqn{G_{55}^< 
=
\frac{|W_{32}W_{34}W_{54}|^2}{|\Delta_c^r|^2}
|g_3|^2|g_4|^2 |g_5|^2 \{ |W_{12}|^2|g_2|^2  (g_1^r v_5^< } \nnm \\
& &+ g_1^<v_5^a) + (g_2^<v_5^a -g_2^a v_5^<)   \} 
\nnm \\
& &
+|W_{34}W_{54}|^2 \frac{ |1-C_{12}^r|^2 }{|\Delta_c^r|^2} 
 |g_4|^2 |g_5|^2 [ g_3^r v_5^<  +g_3^< v_5^a]
 \nnm\\
& &
+|W_{54}|^2\frac{|1-C_{12}^r-C_{32}^r|^2 }{|\Delta_c^r|^2} |g_5|^2 (g_4^< v_5^a -g_4^a v_5^<) 
\nnm\\
& & 
  +\frac{|(1-C_{34}^r)(1-C_{12}^r) -C_{32}^r|^2 }{|\Delta_c^r|^2 }(g_5^r v_5^<+ g_5^< v_5^a). 
\end{eqnarray}
By exchanging "(1,2)" with "(5,4)", we obtain the expression of $G_{11}(\omega)$.

Finally, we input the following relations into the above equations:
\begin{eqnarray}
g_{di}^r (\omega)&=& \frac{a_i(\omega)-i\gamma_i}{D_i(\omega)}, \\
a_i(\omega)&=&\omega-E_{di}-\Lambda_i(\omega), \
D_i =a_i^2 +\gamma_i^2,\nnm \\ \\
g_{di}^< (\omega)&=&iF(\omega)/D_i(\omega), \\
F_i &=& \Gamma_{iL} f_{iL}(\omega)+\Gamma_{iR} f_{iR}(\omega).
\end{eqnarray}
In addition, we symmetrize the current 
$I=(I_L+I_L)/2=(I_L-I_R)/2$ such that
\begin{equation}
v_L^r-v_R^r \Rightarrow \frac{1}{2\pi}(\Gamma^L-\Gamma^R)[P\frac{1}{\omega-E_k }-i\pi \delta(\omega-E_k) ],
\end{equation}
where we assume that $I=I_L=-I_R$.
\begin{widetext}
Then, we can use the equations as follows
\begin{eqnarray}
& &g_1^r v_1^< +g_1^< v_1^a   + (g_1^r v_1^< +g_1^< v_1^a )^* 
\Rightarrow  \frac{1}{D_1(\omega)} \Gamma^R_1 \Gamma^L_1 [f_{1L}(\omega)-f_{1R}(\omega)],  \\
& &g_1^r v_5^< +g_1^< v_5^a + (g_1^r v_5^< +g_1^< v_5^a )^*
\Rightarrow \frac{1}{D_1} \{[\Gamma_{1L}+\Gamma_{1R}] (\Gamma_{5L} f_{5L}(\omega)-\Gamma_{5R} f_{5R}(\omega)) \nnm\\ 
& &-[\Gamma_{5L}-\Gamma_{5R}] (\Gamma_{1L} f_{1L}(\omega)+\Gamma_{1R} f_{1R}(\omega)) \} \delta(\omega-E_{5k}),\\   
& & g_2^<v_5^a -g_2^a v_5^< +(g_2^<v_5^a -g_2^a v_5^<)^*
\Rightarrow 
\pi \{ \Gamma_{5L} [f_{5L}(E_2)-f(E_2)]-\Gamma_{5R} [f_{5R}(E_2)-f(E_2)] \} \delta(E_2-E_{5k}). 
\end{eqnarray}

Thus, the current Eq.(\ref{current5}) is given by calculating from $I=I_{1}+I_{3}+I_{5}$, where 
\begin{equation}
I_i=\frac{e}{\hbar}\sum_{k}\int \frac{d\omega}{2\pi}\{V_k^{(i)}G_{ii}^< +V_{ki}^{*}(G_{ii}^<)^* \}.
\end{equation}
Concretely, we have
\begin{eqnarray}
\lefteqn{I_{3}=
\frac{e}{h}\int d\omega
\Bigl\{ 
 \frac{|1-C_{54}^r|^2|W_{23}|^2|W_{12}|^2}{|\Delta_c^r|^2D_1D_2D_3}
\{[\Gamma_{1L}+\Gamma_{1R}] (\Gamma_{3L} f_{3L}(\omega)-\Gamma_{3R} f_{3R}(\omega))   
 -[\Gamma_{3L}-\Gamma_{3R}] (\Gamma_{1L} f_{1L}(\omega)+\Gamma_{1R} f_{1R}(\omega)) \}  
}\nnm \\
&+&
 \frac{|1-C_{12}^r|^2|W_{43}|^2|W_{54}|^2}{|\Delta_c^r|^2D_3D_4D_5}  
\{[\Gamma_{5L}+\Gamma_{5R}] (\Gamma_{3L} f_{3L}(\omega)-\Gamma_{3R} f_{3R}(\omega)) 
 -[\Gamma_{3L}-\Gamma_{3R}] (\Gamma_{5L} f_{5L}(\omega)+\Gamma_{5R} f_{5R}(\omega)) \}
\nnm \\
&+&
\frac{|1-C_{54}^r|^2|1-C_{12}^r|^2}{|\Delta_c^r|^2D_1(\omega)}
\Gamma^R_3 \Gamma^L_3 [f_{3L}(\omega)-f_{3R}(\omega)] 
\Bigr\}
\nnm \\
 &\Rightarrow & \frac{e}{h}\int d\omega
\Bigl\{ 
 \frac{|1-C_{54}^r|^2|W_{23}|^2|W_{12}|^2}{|\Delta_c^r|^2D_1D_2D_3}
F_{a31}
+\frac{|1-C_{12}^r|^2|W_{43}|^2|W_{54}|^2}{|\Delta_c^r|^2D_3D_4D_5}  
F_{a35}
+\frac{|1-C_{54}^r|^2|1-C_{12}^r|^2}{|\Delta_c^r|^2D_3}
F_{33} 
\Bigr\}.
\end{eqnarray}
The current $I_{1}$ is given by
\begin{eqnarray}
I_{1}
&=& \frac{e}{h} \int \! d\omega 
 \Bigl\{
\frac{|W_{34}W_{32}W_{12}W_{54}|^2}{|\Delta_c^r|^2D_1D_2D_3D_4D_5}
\{[\Gamma_{5L}+\Gamma_{5R}] (\Gamma_{1L} f_{1L}(\omega)-\Gamma_{1R} f_{1R}(\omega)) 
-[\Gamma_{1L}-\Gamma_{1R}] (\Gamma_{5L} f_{5L}(\omega)+\Gamma_{5R} f_{5R}(\omega)) \} 
\nnm \\
& &+\frac{|W_{32}W_{12}|^2  |1-C_{54}^r|^2 }{|\Delta_c^r|^2D_1D_2D_3} 
 \{[\Gamma_{3L}+\Gamma_{3R}] (\Gamma_{1L} f_{1L}(\omega)-\Gamma_{1R} f_{1R}(\omega)) 
-[\Gamma_{1L}-\Gamma_{1R}] (\Gamma_{3L} f_{3L}(\omega)+\Gamma_{3R} f_{3R}(\omega)) \} 
 \nnm\\
& & 
 +\frac{|(1-C_{32}^r)(1-C_{54}^r) -C_{34}^r|^2 }{|\Delta_c^r|^2 D_1}
\Gamma^R_1 \Gamma^L_1 [f_{1L}(\omega)-f_{1R}(\omega)] 
\Bigr\}\nnm\\
&\Rightarrow& \frac{e}{h} \int \! d\omega 
 \Bigl\{
\frac{|W_{34}W_{32}W_{12}W_{54}|^2}{|\Delta_c^r|^2D_1D_2D_3D_4D_5}
F_{b15}
+\frac{|W_{32}W_{12}|^2  |1-C_{54}^r|^2 }{|\Delta_c^r|^2D_1D_2D_3} 
F_{b13} 
+\frac{|(1-C_{32}^r)(1-C_{54}^r) -C_{34}^r|^2 }{|\Delta_c^r|^2 }
 \frac{F_{11}}{D_1}
\Bigr\}.
\end{eqnarray}
$I_5$ is obtained by replacing $1\leftrightarrow 5$ and $2\leftrightarrow 4$. 
We have also defined 
\begin{eqnarray}
F_{a31}&=& [\Gamma_{1L}+\Gamma_{1R}] (\Gamma_{3L} f_{3L}(\omega)-\Gamma_{3R} f_{3R}(\omega))   
 -[\Gamma_{3L}-\Gamma_{3R}] (\Gamma_{1L} f_{1L}(\omega)+\Gamma_{1R} f_{1R}(\omega)), 
\\
F_{a35}&=&
[\Gamma_{5L}+\Gamma_{5R}] (\Gamma_{3L} f_{3L}(\omega)-\Gamma_{3R} f_{3R}(\omega)) 
 -[\Gamma_{3L}-\Gamma_{3R}] (\Gamma_{5L} f_{5L}(\omega)+\Gamma_{5R} f_{5R}(\omega)), 
\\
F_{b15}&=& [\Gamma_{5L}+\Gamma_{5R}] (\Gamma_{1L} f_{1L}(\omega)-\Gamma_{1R} f_{1R}(\omega)) 
-[\Gamma_{1L}-\Gamma_{1R}] (\Gamma_{5L} f_{5L}(\omega)+\Gamma_{5R} f_{5R}(\omega)),  
 \\
F_{b13} &=& 
[\Gamma_{3L}+\Gamma_{3R}] (\Gamma_{1L} f_{1L}(\omega)-\Gamma_{1R} f_{1R}(\omega)) 
-[\Gamma_{1L}-\Gamma_{1R}] (\Gamma_{3L} f_{3L}(\omega)+\Gamma_{3R} f_{3R}(\omega)).
\end{eqnarray}
In the main text, we use
\begin{eqnarray}
F_{ii}&=& \Gamma^R_i \Gamma^L_i [f_{iL}(\omega)-f_{iR}(\omega)], \\
F_{12345}&\equiv& F_{b15}+F_{b51},\\
F_{123}&\equiv& F_{a31}+F_{b13},\\
F_{345}&\equiv& F_{a35}+F_{b53}. 
\end{eqnarray}

\end{widetext}


\end{document}